\documentclass{aa}  
\usepackage{graphicx}
\usepackage{xcolor}

\begin{document} 

\title{Ram pressure stripping in the $z\sim 0.5$ galaxy cluster MS~0451.6-0305}

   \author{F. Durret
          \inst{1}
          \and
          L. Degott
          \inst{1}
          \and
          C. Lobo
          \inst{2,3}
          \and
          H. Ebeling
          \inst{4}
          \and
          M. Jauzac
          \inst{5,6,7,8}
          \and
          S.-I. Tam
          \inst{9}
\thanks{Based on observations made with the NASA/ESA Hubble Space Telescope, obtained from the Space Telescope Science Institute Data Archive, which is operated by the Association of Universities for Research in Astronomy, Inc., under NASA contract NAS 5-26555. These observations are associated with programmes GO-09722, -09836, -10493, and -111591.
This paper has made use of the NASA Extragalactic Database
and of the GAZPAR tool to apply the LePhare software. }  }

\institute{Sorbonne Universit\'e, CNRS, UMR 7095, Institut d'Astrophysique de Paris, 98bis Bd Arago, 75014, Paris, France
 \and
     Instituto de Astrof\'{\i}sica e Ci\^encias do Espa\c co, Universidade do Porto, CAUP, Rua das Estrelas, PT4150-762 Porto, Portugal
            \and
        Departamento de F\'{\i}sica e Astronomia, Faculdade de Ci\^encias, Universidade do Porto, Rua do Campo Alegre 687, PT4169-007 Porto, Portugal
         \and
Institute for Astronomy, University of Hawaii, 2680 Woodlawn Dr, Honolulu, HI 96822, USA         
 \and
Centre for Extragalactic Astronomy, Durham University, South Road, Durham DH1 3LE, UK
    \and
Institute for Computational Cosmology, Durham University, South Road, Durham DH1 3LE, UK
\and 
Astrophysics Research Centre, University of KwaZulu-Natal, Westville Campus, Durban 4041, South Africa
\and
School of Mathematics, Statistics \& Computer Science, University of KwaZulu-Natal, Westville Campus, Durban 4041, South Africa
 \and         
Academia Sinica Institute of Astronomy and Astrophysics (ASIAA), No. 1, Sec. 4, Roosevelt Road, Taipei 10617, Taiwan
  }     
             
\date{}

\authorrunning{Durret et al.}
\titlerunning{Ram pressure stripping galaxies in MS0451}

  \abstract
   {The pressure exerted by the ambient hot X-ray gas on cluster galaxies can lead to the presence of ram pressure stripped (RPS) galaxies, characterized by asymmetric shapes, and, in some cases, tails of blue stars and/or X-ray gas, with 
   increased star formation.}
   {With the aim of increasing the known sample of RPS galaxies at redshift $z\sim 0.5$, notably higher than most RPS samples presently known, we extended our searches for such galaxies to the cluster MS~0451.6-0305.}
{Our study is based on Hubble Space Telescope (HST) imaging in the F814W band (corresponding to a restframe wavelength of 529~nm), covering a region of about $6\times 6$~Mpc$^2$,
an eight magnitude ground-based catalogue with photometric redshifts, and a spectroscopic redshift catalogue.
We defined as cluster members a spectroscopic redshift sample of 359 galaxies within $\pm 4\sigma _v$ of the mean cluster velocity, and a photometric redshift sample covering the [$0.48,0.61$] range. 
We searched for RPS galaxies in the HST images and tested the error on their classification with a Zooniverse collaboration.
We also computed the phase space diagram of RPS candidates in the spectroscopic sample. 
Finally, we ran the LePhare stellar population synthesis code through the GAZPAR interface to analyze and compare the properties of RPS and non-RPS galaxies. 
}
{We find 56 and 273 RPS candidates in the spectroscopic and photometric redshift samples, respectively. They are distributed throughout the cluster 
and tend to avoid high density regions.
The phase space diagram gives the percentages of virialized, backsplash, and infall galaxies. RPS galaxy candidates typically show rather high star formation rates, young ages, and relatively low masses.
We compare all our results to those previously obtained for the massive merging cluster MACS~J0717+3745, at a similar redshift.
}
{This study increases by at least 56 objects if we only consider galaxies with spectroscopic redshifts, and probably much more if galaxies with photometric redshifts are taken into account. This study increases the number of RPS candidates at redshift $z> 0.5$, and
confirms that they host, on average, younger stellar populations and strongly form stars when compared with non-RPS counterparts. The fact that RPS candidates with spectroscopic and with photometric redshifts have comparable properties shows that large samples of such objects could be gathered based on multi-band photometry only, a promising result in view of the very large imaging surveys planned in the coming years (DES, Euclid, LSST, etc.).}


\maketitle

\section{Introduction}

Among the most important mechanisms affecting galaxies in clusters are ram pressure stripping \citep{GunnGott72},
hereafter RPS, which affects the gas contained in galaxies, and tidal effects {\citep[e.g.][]{Toomre72} which affect both gas and stars and can lead to harrassment \citep{Moore+96} when high-speed encounters predominate (as is the case in massive clusters). A complete summary of all the physical processes affecting galaxy evolution and taking place in clusters can be found,  for example, in the introduction of the paper by \citet{Poggianti+17}. 

Ram Pressure Stripped galaxies, which are the topic of the present paper, can be signalled by asymmetric shapes, sharp edges, blue knots, and/or tails, and they often show enhanced star formation. Not only are they interesting objects by themselves, allowing us to study galaxy evolution in dense environments, but they also provide  important indications for the properties of the clusters in which they are embedded. However, even though we can learn much from these objects, they are still difficult to identify.

While detailed hydrodynamical simulations have only recently started to be developed for RPS galaxies \citep{Yun+19}, observational evidence for gas stripping at various wavelengths has been accumulating for a number of years. First evidence came from the HI deficiency detected in cluster galaxies \citep{Haynes+84}, particularly in the well-known examples of Virgo \citep{Cayatte+90} and Coma \citep{Bravo+01}. 
Although the star formation in a galaxy deprived of some of its gas is expected to eventually drop,
the observation of ionized gas at optical wavelengths, a good tracer of star formation, shows that in some cluster galaxies, likely caught at an earlier phase of action of this mechanism, the star formation rate (SFR) can increase for a limited amount of time. This is attributed to the compression effect of ram pressure, which can also be at the origin of spectacular tails of stripped, ionized gas reaching tens of kpc in length, and often containing blue knots revealing intense star formation. 
Many observations of such ionized tails are now available \citep[][among others]{Owen+06,Cortese+07,Sun+07,Yoshida+08,Smith+10,Yagi+10,Fossati12,Boselli16} and galaxies with such tails of star-forming material, likely produced by RPS, were named ''jellyfish galaxies'' following \cite{Bekki09}. 
A systematic search for jellyfish galaxies by \cite{Poggianti+16} in the WINGS and OMEGAWINGS surveys showed that
disturbed morphologies suggestive of stripping phenomena are ubiquitous in clusters and could also be present in
groups and low mass halos. RPS can also have more modest effects on galaxies, leading to features such as asymmetric shapes or sharp edges on one side of the galaxy, that can indicate in which direction the galaxy is moving relatively to the intracluster 
medium \citep[see e.g.][]{McPartland+16}.

Hereafter, we use the generic term of RPS galaxies, since not all RPS galaxies look like jellyfish, and also because the term jellyfish refers to very specific signatures. To illustrate this, we refer the reader to a conspicuous example of a ram pressure stripped galaxy with an H$\alpha$ (and X-ray) tail that was found by \citet{Laudari+22} to show little star formation activity. This is not an isolated case \citep[see references in][]{Laudari+22} but these authors use this argument to suggest that this particular object, and others alike, cannot be classified as a jellyfish. This finding also exposes the complexity of this phenomenon that, although fascinating, involves a lengthy discussion that is beyond the scope of this paper.
\cite{Vulcani+21} and \citet{Laudari+22} further stressed the difficulty of selecting ``clean'' samples of RPS galaxies, a problem that we briefly discuss in Sections~\ref{sec:concerns} and \ref{sec:discu}. 

VLT/MUSE has been very successfully used to observe RPS galaxies, first by \citet{Fumagalli+14}, who detected a double tail reaching 80 kpc, previously seen in X-rays, in ESO~137-001 (z=0.01625). A complementary study with APEX \citep{Jachym+14} uncovered an exceptionally long molecular tail in this galaxy, and follow-up observations with ALMA \citep{Jachym+19} allowed the first detection of the molecular gas at the heads of several fireballs located in the complex tail structure of this spectacular galaxy. This confirmed the importance of multi-wavelength observations to fully understand these objects.
Radio observations of 20 Coma galaxies were then obtained by \cite{Chen+20}, showing that radio continuum emission could also be detected in ionized gas tails. 

A dedicated, programme using VLT/MUSE was pursued with the GASP(GAs Stripping Phenomena in galaxies with MUSE) survey of  nearby RPS galaxies \citep{Poggianti+17}, leading to a large number of very interesting results presented in more than 30 papers. 
VLT/MUSE data on GASP were also complemented with data at other wavelengths such as ALMA, VLA, UVIT, and {\it Chandra} \citep{Poggianti+19}, the latter paper proposing that heating due to interaction with the intracluster medium is responsible for the X-ray tail. 

The molecular gas in RPS galaxies was analyzed in JW100 \citep{Moretti+20b} and four other comparable objects \citep{Moretti+20a}. They interpreted this large molecular gas content as newly born from stripped HI gas or recently condensed from stripped diﬀuse molecular gas. 
\cite{Ramatsoku+20} observed two RPS galaxies of the GASP survey in HI and H$\alpha$  and found them to be similar, based on their H I content, stellar mass, and star formation rate, an unexpected finding in view of their different environments. 
\cite{Deb+20} observed in HI one of the most spectacular GASP galaxies, JO204, and detected an extended 90~kpc long ram-pressure stripped tail of neutral gas, stretching beyond the 30 kpc long ionized gas
tail and pointing away from the cluster centre. The comparison of the distributions and kinematics of the neutral and ionized gas in the tail indicates a highly turbulent medium.

In four RPS galaxies, \cite{Moretti+20b} found within the disk an H$_2$ molecular gas mass a factor
between 4 and $\sim 100$ times higher than the neutral gas mass, and interpreted these results as due to the fact that ram pressure in disks of galaxies during the RPS phase causes a very efficient conversion of HI into H$_2$.

Recently, still another result of the GASP survey, \cite{Vulcani+21} separated galaxies into several categories, depending on the different processes they were undergoing  in low-density environments: galaxy–galaxy interactions, mergers, ram pressure stripping, cosmic web stripping, cosmic web enhancement, gas accretion, and starvation. 

Besides the GASP survey, a few other authors studied large samples of RPS galaxies at various wavelengths. 
\cite{Roberts+21} observed galaxies with LOFAR in a large sample of $\sim 500$ galaxy groups and clusters, and showed that galaxies also undergo RPS in groups, with differences between groups and clusters. They underline in particular that: 
the frequency of RPS galaxies is highest in clusters and lowest in low-mass groups; there is weaker ram pressure stripping in groups relative to clusters; many jellyfish galaxies in groups are consistent with having already passed pericentre, which does not seem to be the case for jellyfish galaxies in clusters; and unlike jellyfish galaxies in clusters, jellyfish galaxies in groups do not have systematically enhanced SFR. 

The study of RPS galaxies at intermediate redshift ($z=0.3-0.5$)  started quite recently. 
Based on the slitless spectroscopy taken as part of the Grism Lens-Amplified Survey from Space (GLASS), 
\cite{Vulcani+16} presented an extended analysis of the spatial distribution of star formation in 76 galaxies in 10 clusters at $0.3<z<0.7$, together with 85 foreground and background galaxies in the same redshift range as a field sample. They find a large diversity of H$\alpha$ morphologies, suggesting a diversity of physical processes: some  galaxies are regular (30\%), some are asymmetric or jellyfish (28\%), and their shape can be attributed to RPS, which is much less prominent in the field (2\%). 
In a following paper based on the same data, \cite{Vulcani+17} correlate the properties of the H$\alpha$ emitters to several tracers of the cluster environment and conclude that local effects, uncorrelated to the
clustercentric radius, play a more important role than the global environment in shaping galaxy properties.

VLT/MUSE also led to new results on RPS galaxies in intermediate redshift clusters. \cite{Boselli+19} observed with VLT/MUSE a cluster at z = 0.73 containing two massive star-forming galaxies with extended tails of diffuse ionized gas reaching a projected length of $\sim$100~kpc and not associated with any stellar component. 
\cite{Moretti+22} found 13 RPS galaxies in the clusters A2744 (z=0.306) and A370 (z=0.373) and detected not only H$\alpha$ but [OII]$\lambda\lambda 3726,3729$. They found that the resolved [OII]/H$\alpha$ line ratio in the stripped tails is exceptionally high compared to that in the disks of these galaxies, suggesting
lower gas densities and/or an interaction with the hot surrounding ICM. With the IFU device of the Keck telescope, \cite{KalitaEbeling19} observed the A1758N\_JFG1 jellyfish galaxy at z=0.28 and detected [OII]3727 emission as far as 40~kpc from the cluster centre.

Though dedicated IFU studies are extremely rich and relevant for understanding the details of the evolution of galaxies subject to RPS and of the direct consequences for their star formation and metallicity histories, larger samples are still lacking, in particular at intermediate redshifts. These can be used to tentatively infer some statistical trends and serve as a starting point for follow-up studies.

With this in mind, a search for RPS candidates has been undertaken in the HST images of 40 DAFT/FADA and CLASH surveys by \citet{Durret+21} in the redshift range $0.2<z<0.9$. They studied in particular the case of MACS~J0717.5+3745 (hereafter MACS0717), a very interesting X-ray luminous and complex merging cluster located at z=0.5458. For this system, not only HST data were available for the cluster and its extended filament, but also a spectroscopic redshift catalogue with several hundreds of objects, and ground-based magnitudes in six optical and two infrared photometric bands in a large region around the cluster.}

Merging clusters should actually be sweet spots for searches of RPS galaxies, as suggested by different authors \citep[e.g.][]{Owers+12,McPartland+16,KalitaEbeling19,EbelingKalita19,Moretti+22}, so additional candidates in the sample of \citet{Durret+21} are worth further  exploration.

With the aim of increasing the sample of RPS galaxies at z $\sim0.5$ and comparing the two systems, we present here observations of another cluster, MS~0451.6-0305 (hereafter MS0451, also named MACS~J0454.1-0300), located at redshift $z=0.5377$, very similar to that of MACS0717. For this cluster, we have data comparable to those for MACS0717, namely large spatial coverage up to  clustercentric radius larger than $r_{100}$ (though we have fewer redshifts for MS0451). MS0451 was identified as a post-merger system \citep{Tam+20}, and is a member of the DAFT/FADA sample (massive clusters with HST data in the $0.4<z<0.9$ redshift range). Its galaxy density map shows no strong elongation or particular features, but it was noted that the peak of the density map (where the contours reach a maximum) does not coincide with the position of the cluster given by NED, the displacement
being about 0.011 deg, corresponding to 0.25 Mpc at the cluster distance. The weak lensing analysis detects two peaks \citep{Martinet+16}, following the general elongation of the galaxy density map of \citet{Durret+16}.
We present in this paper a search for RPS candidates, selected as cluster members on the basis of spectroscopic and photometric redshifts.

We assume here H$_0=70$~km~s$^{-1}$~Mpc$^{-1}$, $\Omega_\Lambda=0.7$ and $\Omega _{\rm m}=0.3$. At this redshift, the scale is 6.34~kpc/arcsec. Magnitudes are given in the AB system.
The cluster centre is taken from NED: RA(J2000)=73.54635~deg, DEC(J2000)=$-3.01494$~deg.

\section{The data}

Our study is based on a large HST/ACS mosaic in the F814W filter covering a region of about 6 × 6 Mpc$^2$ at the cluster redshift, collected for the programmes GO$-09836$, $-10493$, and $-11591$. We also examined F555W imaging from a single ACS pointing obtained for GO-09722 but found the covered area too small to be significant for the purpose of this work. 
We reduced the imaging data with the pyHST software 
package\footnote{https://github.com/davidharvey1986/pyHST}. For the details of data reduction process, we refer the reader to \cite{Tam+20}.

\subsection{RPS characterization: systematic concerns}
\label{sec:concerns}
Since the most easily recognizable RPS features involve triggered star formation, either at the ICM/ISM interface or in the debris trail (see \cite{Cortese+07} or \cite{KalitaEbeling19} for textbook examples), images in a band bluer than the 4000~\AA\ break at the cluster redshift are best suited to identify and characterize RPS galaxies. At the redshift of our target, such an endeavour would thus require wide-area imaging in F555W or an even bluer HST/ACS filter, which are unfortunately not available for MS0451. We nonetheless attempt to identify RPS features in the F814W filter, relying less on ``jellyfish" tentacles but basing our selection on broader morphological features such as global asymmetry,  sharp edges and/or notable unilateral distortion of disks or spiral arms. Although we have excluded galaxies with a nearby galaxy neighbour (see next section), the resulting sample of RPS candidates will invariably be contaminated by minor mergers in addition to irregular galaxies and dusty spirals. We estimate by eye inspection that at the very most 15 galaxies in our sample of 56 spectroscopic RPS candidates could be minor mergers (namely \#4, \#6, \#7, \#11, \#15, \#16, \#18, \#24, \#30, \#32, \#36, \#44, \#45, \#46, \#49). 
We have checked that these 15 galaxies are not located in a specific zone of Fig.~11 (13 out of 15 are in the virialized zone) and of Fig.~16 (they cover the whole mass and SFR ranges).
On the other hand, we may have missed RPS galaxies with features emitting more in the blue, due to the red filter we used. We keep these {\it caveats} in mind in our analysis. Acknowledging the importance (and lack) of imaging in a bluer filter, we did not attempt to assign a velocity vector (relative to the ICM) to each candidate. 

We can note that \cite{McPartland+16} and \cite{Roman+21} suggested and tested various automatic classification algorithms, but since these techniques are still under development we preferred not to use them.

\subsection{Sample 1 (spectroscopic redshifts)}

We put together the spectroscopic redshift catalogue of \citet{Tam+20} and the redshifts taken from NED in the area covered by the HST mosaic, keeping in priority the values from \citet{Tam+20} when they were also present in NED. For the 129 galaxies with redshifts in the cluster range and appearing both in NED and in \citet{Tam+20}, the redshifts agree very well: the mean value of the difference between the redshifts is $4.1\times 10^{-5}$ with a dispersion of $3.3\times 10^{-5}$.
In order not to miss high velocity galaxies, we chose to consider as cluster members all the galaxies with velocities within $\pm 4\sigma _v$ of the mean cluster velocity (see section \ref{sec:phase}). This corresponds to the redshift interval
$0.5086<z<0.5678$. Within this range, we finally obtained a spectroscopic redshift catalogue of 359 galaxies. 

Two of us (FD and LD) looked at the HST images of all these galaxies and gave them a J classification (respectively J$_F$ and J$_L$ for FD and LD), following \citet{Poggianti+16}. This classification grades galaxies from J=1 to J=5, with the value of J increasing as the likeliness of being affected by RPS increases.
We chose as final classification the mean of the two values, which usually did not differ much, J$_{mean}$. We eliminated objects with J$_{mean}<1$ (i.e., considered as non-RPS by one of us and classified as J=1 by the other) and thus obtained the list of 56 RPS candidates presented in Table~\ref{tab:jelly}. We note that in this process we also eliminated galaxies that could be undergoing a gravitational interaction with a neighbouring galaxy (since their shape would then not be due to ram pressure stripping) or that had a shape similar to that of a gravitational arc. 

In order to measure the F814W magnitudes, we computed the zero points ${\rm ZP_{AB}}$ by applying the {\it HST} formula:\footnote{http://www.stsci.edu/hst/instrumentation/acs/data-analysis/zeropoints}
\small
$${\rm -2.5*log10(PHOTFLAM)-5*log10(PHOTPLAM)-2.408},$$
\normalsize
\noindent
where the PHOTFLAM and PHOTPLAM values were found in the image headers (they were identical for all the images, each having an
exposure time of 2036~s). We then ran SExtractor \citep{Bertin96} on the mosaic 
and retrieved the MAG\_AUTO magnitudes. The values given in Table~\ref{tab:jelly} are not corrected for Galactic extinction.

\begin{table*}
\centering
\caption{Sample 1 RPS candidates in MS0451 (galaxies with spectroscopic redshifts). The columns are: running number, J2000 right ascension and declination (in degrees), spectroscopic redshift, flag for the origin of the redshift (1 for \citet{Tam+20} and 2 for NED), J classifications J$_L$ and J$_F$ according to two of us (LD and FD), mean J classification, F814W magnitude and its error, distance to cluster centre in units of r$_{200}$, $\Delta v/\sigma _v$ (difference between the galaxy and the cluster velocities divided by the cluster velocity dispersion).}

\begin{tabular}{rrrrrrrrrrrr}
\hline \hline
Number & RA & DEC & z & flag & J$_L$ & J$_F$ & J$_{mean}$ & F814W & error & r/r$_{200}$ & $\Delta v/\sigma _v$ \\ 
\hline
 1 & 73.3313 & -3.0023 & 0.5390 & 2 & 5 & 4 & 4.5 & 20.792 &  0.004 & 2.381 & 0.176\\ 
 2 & 73.4129 & -2.9304 & 0.5342 & 1 & 3 & 3 & 3.0 & 22.407 &  0.011 & 1.746 &-0.474\\ 
 3 & 73.4182 & -3.0043 & 0.5350 & 2 & 4 & 3 & 3.5 & 23.149 &  0.014 & 1.425 &-0.365\\ 
 4 & 73.4277 & -3.0033 & 0.5343 & 1 & 2 & 2 & 2.0 & 22.828 &  0.009 & 1.318 &-0.460\\ 
 5 & 73.4304 & -3.0367 & 0.5336 & 1 & 1 & 2 & 1.5 & 20.731 &  0.003 & 1.302 &-0.556\\ 
 6 & 73.4361 & -3.0897 & 0.5460 & 2 & 3 & 3 & 3.0 & 21.384 &  0.004 & 1.479 & 1.118\\ 
 7 & 73.4557 & -2.9902 & 0.5488 & 1 & 1 & 1 & 1.0 & 21.932 &  0.005 & 1.035 & 1.492\\  
 8 & 73.4600 & -3.1263 & 0.5481 & 2 & 5 & 5 & 5.0 & 22.164 &  0.005 & 1.563 & 1.399\\ 
 9 & 73.4731 & -2.9390 & 0.5405 & 1 & 3 & 3 & 3.0 & 21.092 &  0.003 & 1.168 & 0.379\\ 
10 & 73.4742 & -3.1007 & 0.5416 & 2 & 5 & 5 & 5.0 & 21.903 &  0.008 & 1.236 & 0.527\\ 
11 & 73.4748 & -2.9576 & 0.5320 & 2 & 2 & 3 & 2.5 & 22.796 &  0.012 & 1.014 &-0.773\\ 
12 & 73.4777 & -3.0037 & 0.5300 & 2 & 2 & 2 & 2.0 & 22.751 &  0.009 & 0.764 &-1.046\\  
13 & 73.4911 & -3.1496 & 0.5440 & 2 & 2 & 2 & 2.0 & 23.062 &  0.012 & 1.615 & 0.850\\  
14 & 73.4931 & -2.9430 & 0.5450 & 2 & 1 & 1 & 1.0 & 21.622 &  0.004 & 0.991 & 0.984\\ 
15 & 73.4933 & -2.9684 & 0.5278 & 1 & 2 & 3 & 2.5 & 22.715 &  0.007 & 0.781 &-1.347\\ 
16 & 73.4947 & -2.9820 & 0.5381 & 1 & 2 & 2 & 2.0 & 21.656 &  0.005 & 0.672 & 0.055\\ 
17 & 73.4946 & -2.9827 & 0.5390 & 2 & 2 & 1 & 1.5 & 21.756 &  0.006 & 0.676 & 0.176\\ 
18 & 73.4979 & -2.9677 & 0.5512 & 1 & 1 & 1 & 1.0 & 21.183 &  0.005 & 0.748 & 1.812\\ 
19 & 73.4991 & -3.0529 & 0.5348 & 1 & 3 & 4 & 3.5 & 21.618 &  0.005 & 0.669 &-0.393\\ 
20 & 73.5044 & -3.0522 & 0.5405 & 1 & 2 & 2 & 2.0 & 21.326 &  0.003 & 0.620 & 0.379\\ 
21 & 73.5062 & -2.9934 & 0.5313 & 1 & 3 & 3 & 3.0 & 22.833 &  0.011 & 0.502 &-0.869\\ 
22 & 73.5069 & -2.9466 & 0.5404 & 1 & 1 & 2 & 1.5 & 22.230 &  0.007 & 0.875 & 0.365\\ 
23 & 73.5078 & -2.9789 & 0.5288 & 1 & 1 & 1 & 1.0 & 22.634 &  0.007 & 0.583 &-1.210\\ 
24 & 73.5138 & -2.9891 & 0.5310 & 2 & 1 & 1 & 1.0 & 22.581 &  0.007 & 0.457 &-0.909\\ 
25 & 73.5217 & -2.9974 & 0.5388 & 2 & 3 & 3 & 3.0 & 21.429 &  0.003 & 0.334 & 0.149\\ 
26 & 73.5288 & -2.9073 & 0.5357 & 2 & 1 & 2 & 1.5 & 23.393 &  0.010 & 1.209 &-0.270\\ 
27 & 73.5311 & -2.9636 & 0.5330 & 2 & 5 & 5 & 5.0 & 22.011 &  0.006 & 0.593 &-0.637\\ 
28 & 73.5431 & -3.1726 & 0.5463 & 2 & 2 & 3 & 2.5 & 21.986 &  0.005 & 1.751 & 1.158\\ 
29 & 73.5507 & -3.1334 & 0.5487 & 2 & 1 & 2 & 1.5 & 22.319 &  0.008 & 1.316 & 1.479\\ 
30 & 73.5525 & -3.0554 & 0.5524 & 2 & 2 & 3 & 2.5 & 20.309 &  0.003 & 0.457 & 1.972\\ 
31 & 73.5553 & -3.0057 & 0.5571 & 2 & 1 & 1 & 1.0 & 24.271 &  0.022 & 0.146 & 2.595\\ 
32 & 73.5560 & -3.0301 & 0.5478 & 2 & 3 & 3 & 3.0 & 21.979 &  0.006 & 0.201 & 1.359\\ 
33 & 73.5593 & -3.0662 & 0.5310 & 2 & 1 & 1 & 1.0 & 23.007 &  0.012 & 0.590 &-0.909\\ 
34 & 73.5594 & -3.0255 & 0.5404 & 1 & 1 & 1 & 1.0 & 21.011 &  0.002 & 0.188 & 0.365\\ 
35 & 73.5625 & -3.0062 & 0.5240 & 2 & 5 & 5 & 5.0 & 19.979 &  0.001 & 0.206 &-1.868\\ 
36 & 73.5631 & -3.0599 & 0.5350 & 2 & 1 & 1 & 1.0 & 22.169 &  0.004 & 0.524 &-0.365\\ 
37 & 73.5640 & -3.1474 & 0.5420 & 2 & 2 & 2 & 2.0 & 22.072 &  0.007 & 1.484 & 0.581\\ 
38 & 73.5716 & -3.0991 & 0.5441 & 2 & 5 & 5 & 5.0 & 20.767 &  0.004 & 0.978 & 0.863\\ 
39 & 73.5743 & -3.1364 & 0.5471 & 2 & 2 & 2 & 2.0 & 20.083 &  0.002 & 1.382 & 1.265\\ 
40 & 73.5823 & -3.0678 & 0.5446 & 2 & 3 & 3 & 3.0 & 21.826 &  0.004 & 0.712 & 0.930\\ 
41 & 73.5830 & -3.0264 & 0.5472 & 2 & 2 & 2 & 2.0 & 22.498 &  0.007 & 0.429 & 1.278\\ 
42 & 73.5885 & -3.0571 & 0.5420 & 2 & 2 & 2 & 2.0 & 22.773 &  0.008 & 0.663 & 0.581\\ 
43 & 73.5968 & -3.0625 & 0.5410 & 2 & 1 & 1 & 1.0 & 22.977 &  0.010 & 0.772 & 0.446\\ 
44 & 73.5974 & -3.0635 & 0.5420 & 2 & 1 & 1 & 1.0 & 21.059 &  0.003 & 0.784 & 0.581\\ 
45 & 73.5996 & -3.1177 & 0.5350 & 2 & 1 & 1 & 1.0 & 23.277 &  0.011 & 1.282 &-0.365\\
46 & 73.6035 & -3.1313 & 0.5340 & 2 & 1 & 2 & 1.5 & 21.775 &  0.005 & 1.437 &-0.501\\ 
47 & 73.6274 & -3.1215 & 0.5391 & 2 & 1 & 2 & 1.5 & 21.614 &  0.004 & 1.484 & 0.190\\ 
48 & 73.6317 & -3.1075 & 0.5490 & 2 & 3 & 3 & 3.0 & 20.989 &  0.003 & 1.397 & 1.519\\ 
49 & 73.6349 & -3.0398 & 0.5503 & 1 & 1 & 1 & 1.0 & 22.086 &  0.006 & 1.019 & 1.692\\ 
50 & 73.6433 & -3.1345 & 0.5400 & 2 & 1 & 1 & 1.0 & 22.317 &  0.008 & 1.708 & 0.311\\ 
51 & 73.6461 & -3.1212 & 0.5367 & 2 & 1 & 1 & 1.0 & 22.783 &  0.007 & 1.616 &-0.135\\ 
52 & 73.6481 & -3.1027 & 0.5379 & 2 & 1 & 1 & 1.0 & 21.462 &  0.004 & 1.489 & 0.027\\ 
53 & 73.6542 & -3.1415 & 0.5400 & 2 & 2 & 2 & 2.0 & 21.138 &  0.003 & 1.845 & 0.311\\ 
54 & 73.6650 & -3.1494 & 0.5390 & 2 & 3 & 3 & 3.0 & 21.427 &  0.004 & 1.990 & 0.176\\ 
55 & 73.6744 & -3.1331 & 0.5450 & 2 & 1 & 2 & 1.5 & 22.957 &  0.007 & 1.930 & 0.984\\ 
56 & 73.6819 & -3.0045 & 0.5470 & 2 & 2 & 2 & 2.0 & 22.868 &  0.009 & 1.507 & 1.252\\ 
\hline
\end{tabular}
\label{tab:jelly}
\end{table*}

\begin{figure}
\begin{center}
\includegraphics[width=0.3\textwidth]{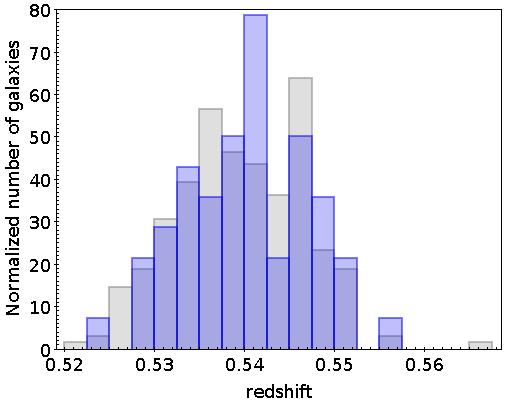}
\end{center}
\caption{Spectroscopic redshift histograms for 
galaxies in the [0.5086,0.5678] redshift interval, normalized to the same area. The grey histogram shows non-RPS galaxies and the blue one the 56 RPS candidates (Sample~1).}
\label{fig:histoz}
\end{figure}

\begin{figure}
\begin{center}
\includegraphics[width=0.3\textwidth]{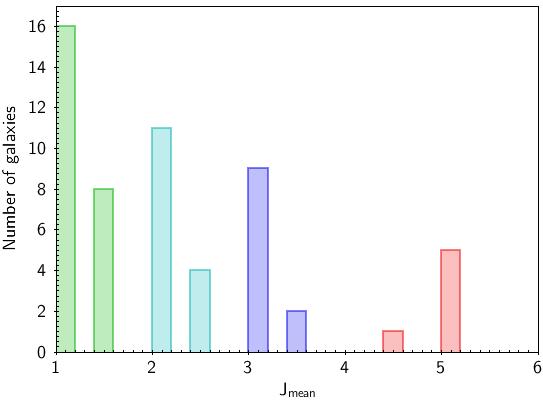}
\end{center}
\caption{Histogram of the mean J classification J$_{mean}$ for Sample~1. The colours correspond to the following figures for sub-samples 1a (green), 1b (cyan), 1c (blue), and 1d (red).}
\label{fig:histoJ}
\end{figure}

\begin{figure}
\begin{center}
\includegraphics[width=0.3\textwidth]{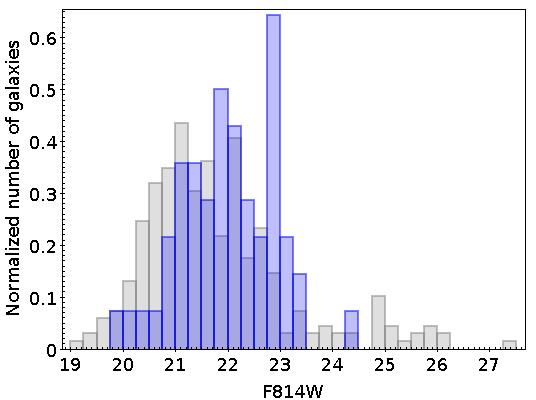}
\end{center}
\caption{Magnitude histograms of the galaxies in the [0.5086,0.5678] redshift interval in the F814W filter, normalized to the same area. The grey histogram shows non-RPS galaxies and the blue one the 56 RPS candidates (Sample~1).}
\label{fig:histomag}
\end{figure}

The 56 RPS candidates with spectroscopic redshifts in the cluster range (hereafter Sample~1) are listed in Table~\ref{tab:jelly} and the corresponding individual images are displayed in the Appendix.
They are in the magnitude range $19.98 \leq F814W \leq 24.27$.

The histograms of the redshifts, mean J values and F814W magnitudes are shown in Figs.~\ref{fig:histoz}, \ref{fig:histoJ} and \ref{fig:histomag} respectively, for the 359 galaxies with spectroscopic redshifts in the cluster range, among which the 56 RPS candidates. We can see from
Fig.~\ref{fig:histoz} that the redshift distributions of the overall cluster population and of the RPS candidates are quite similar. 
The median redshifts of the two samples are 0.5394 and 0.5400 respectively, and a Kolmogorov-Smirnov (KS) test does not reject the hypothesis that these two samples are drawn from the same parent population (with a probability of 12\%).

Fig.~\ref{fig:histoJ} shows the distribution of J values. Based on this figure, the bins we consider hereafter are as follows: J$_{mean}$=1 and 1.5 (24 galaxies, Sample 1a), 
J$_{mean}$=2 and 2.5 (15 galaxies, Sample 1b), 
J$_{mean}$=3 and 3.5 (11 galaxies, Sample 1c), and
J$_{mean}$=4 to 5 (6 galaxies, Sample 1d). The results in these four bins will be discussed in the following sections.
We adopt the following colour code for the sub-samples: 1a in green, 1b  in cyan, 1c in blue, and 1d in red.

The median F814W magnitudes of the two spectroscopic samples 
(the 359 galaxies in the cluster redshift range and the 56 RPS galaxies)
are 21.517 and 21.986 respectively, suggesting that RPS galaxies tend to be fainter than the overall sample. A KS test does not reject the hypothesis that these two samples  are drawn from the same parent population at a probability of 24\%.

\subsection{Test on the J classification}
\label{sec:zoo}

\begin{figure}
\begin{center}
\includegraphics[width=0.3\textwidth]{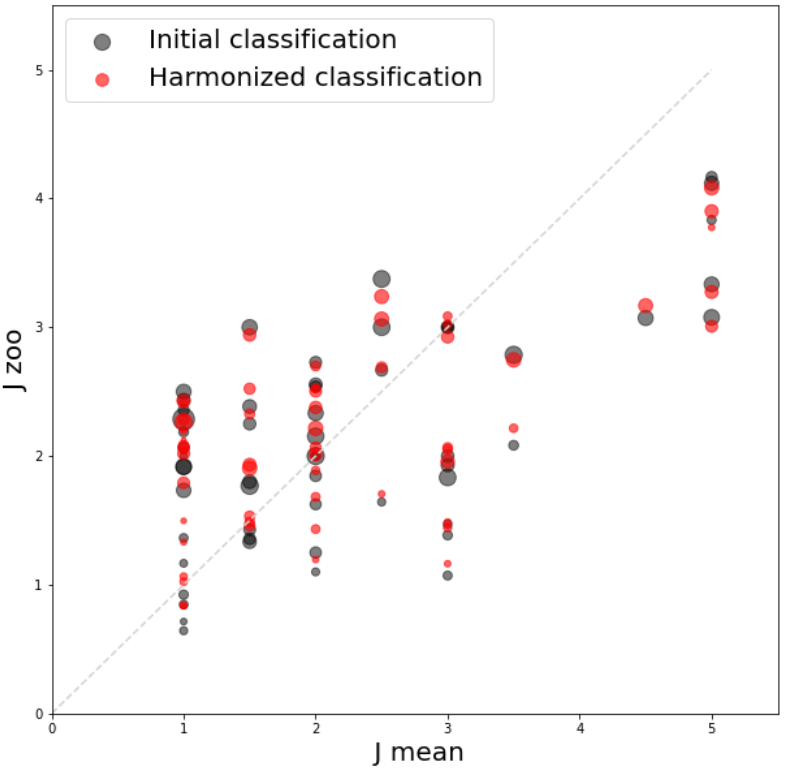}
\end{center}
\caption{Zooniverse J classification as a function of our mean classification for the 56 RPS candidates of Sample~1. The grey points show the initial classification and the red ones that obtained after harmonizing the classifications. The point sizes increase with the standard deviation on the measurement of J.
}
\label{fig:zoo}
\end{figure}

In order to test the reliability of our J classification, we used the Zooniverse software\footnote{https://www.zooniverse.org/} 
and asked  volunteers (mostly graduate students in astronomy with or without knowledge on galaxies) to classify part or all of the 56 RPS candidates of Sample~1. As a training, we first gave them two examples of each class (only one for class 4). In total, 23 people participated, but 10 of them classified fewer than 10 galaxies, so their results were not taken into account, and some galaxies were classified by the same person several times.

Each galaxy taken into account here was classified between 11 and 18 times. 
For the 56 galaxies of Sample~1, Table~\ref{tab:zoo} gives the Zooniverse results: the number of times the galaxy was classified, the minimum and maximum values of J, the mean value of J on all the measurements and the corresponding dispersion.
This dispersion varies between 0.61 and 1.98. 

Another bias could be due to the fact that some of the volunteers tend to grade too optimistically or pessimistically compared to others, and a weighing of the J values should then be applied. To quantify this rigorously, we would need each person to have classified all 56 galaxies, and this was not the case. However, we have tried to harmonize the J grades given by each classifier. The set of grades given by each volunteer have a mean value $mean_{ref}$ and a standard deviation $\sigma_{ref}$. 
For this harmonization, we used the following relation: 
$$J_{i,harmonized}=(\sigma_{ref}/\sigma)*(J_{i}-mean)+mean_{ref}, $$
\noindent
where classifier $i$ gave a series of grades $J_i$, that have a mean value, $mean$, and a standard deviation, $\sigma$. 
It then becomes possible to compare the different classifications.

Figure~\ref{fig:zoo} shows that galaxies rated with a small standard deviation tend to be correlated with low marks, and that the harmonization increases their values. High grade galaxies have larger standard
deviations. One can conclude that volunteers have, most of the time, given lower grades compared to us, and were reluctant to give high grades (especially J=4 or 5).

We can see from Fig.~\ref{fig:zoo} that the difference between the classification obtained with the Zooniverse and ours is between 1 and 2, except for J=5 where it becomes smaller. This confirms our general feeling that, even with some training, classifiers often hesitate between two consecutive values of J.

Interestingly, we find no correlation between the dispersion in the J measurements from the Zooniverse and the galaxy magnitude, so the classification does not appear more secure for bright objects than for faint ones.

As pointed by the referee, when many people classify the same galaxy and the mean value is considered, chances of getting a J=5 mean value are very low (all classifiers must give a value of 5), and these chances decrease with the increasing number of classifiers. In contrast, to get an average value of three, for example, many more combinations of classifications can be given. This bias is difficult to quantify and to correct but should be kept in mind. 

This classification method could contribute to boosting the development of machine learning algorithms with satisfactory performances by increasing the size of the learning sample, thus providing a still-lacking precise tool of shape identification to classify RPS galaxies. However, we must remember how difficult it is to select RPS galaxies in an unambiguous way.

\subsection{Sample~2 (photometric redshifts)}

We have an 8-band magnitude catalogue (in the U, B, V, R$_c$, I$_c$, Z, J, and K bands) covering a larger zone than the HST mosaic and containing 14731 photometric redshifts (see \citet{Tam+20})
that we consider here to make a tentative study of a large sample of RPS galaxy candidates in this cluster. 
If we consider that the typical error on photometric redshifts is $0.05(1+z)$, the redshift interval for galaxies to be possible cluster members is approximately $0.48<z<0.61$. This corresponds to about $\pm 8.7\sigma _v$, a large value that is due to the large error on photometric redshifts.
There are 2180 galaxies in this photometric redshift interval, out of which 1326 are inside the HST mosaic.

We first checked the general agreement between the classifications made by two of us (FD and LD) for Sample 1 by calculating the mean (standard deviation) of J$_F$ and J$_L$. We find respective values of 2.125 (1.27) and
2.286 (1.19), showing a good agreement between the two classifiers. Therefore, only one of us (LD) looked at the HST images of the 1326 galaxies with photometric redshifts in the $0.48<z<0.61$ redshift interval and gave them a J classification, resulting in 217 RPS candidates.
 The catalogue of 217 RPS candidates based on their photometric redshift will be available in electronic form at the CDS via anonymous ftp to cdsarc.u-strasbg.fr (130.79.128.5)
or via http://cdsweb.u-strasbg.fr/cgi-bin/qcat?J/A+A/. This catalogue contains the following columns: number, RA, DEC, photometric redshift, J classification, F814W magnitude and its error. 
We added to this sample the 56 galaxies of Sample~1, which all have a photometric redshift in the $0.48<z<0.61$ interval, and thus created a sample with 273 RPS candidates, that we refer to as Sample~2 hereafter.

Obviously, some of these galaxies do not belong to the cluster. 
To have an idea on the errors caused by a selection based  on photometric redshifts, we can 
analyze several facts. First, we note that all galaxies of Sample~1 pass the photometric redshift selection and are thus included in Sample~2. 
Second, we have a full spectroscopic redshift catalogue of 620 galaxies, out of which 359 and 261 have spectroscopic redshifts respectively in and out of the cluster [0.5086,0.5678] spectroscopic range. Out of the 261 galaxies that are out of the cluster spectroscopic range, 27 have a photometric redshift in the [0.48,0.61] photometric redshift cluster range, or 10\%. 
If we now consider the 273 RPS candidates of Sample~2, which are all cluster members based on their photometric redshift, there are 8  galaxies with  spectroscopic redshifts outside the cluster spectroscopic range. 

Finally, even though the morphological properties of these RPS candidates - that led to their selection - are interpreted as due to their interaction with the dense intracluster medium, some contaminants belonging to foreground/background structures are still likely to be present. This is discussed in Sect.~\ref{sec:discu}.

\begin{figure}
\begin{center}
\includegraphics[width=0.3\textwidth]{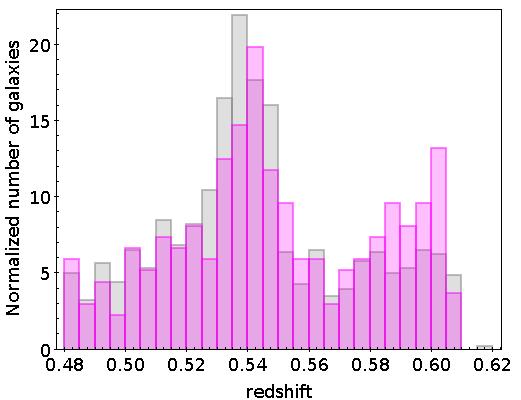}
\end{center}
\caption{Photometric redshift histograms normalized to the same area: the 1326 galaxies in the [0.48,0.61] redshift interval are shown in grey and the 273 RPS candidates of Sample~2 in magenta.}
\label{fig:histo_zphot}
\end{figure}

\begin{figure}
\begin{center}
\includegraphics[width=0.3\textwidth]{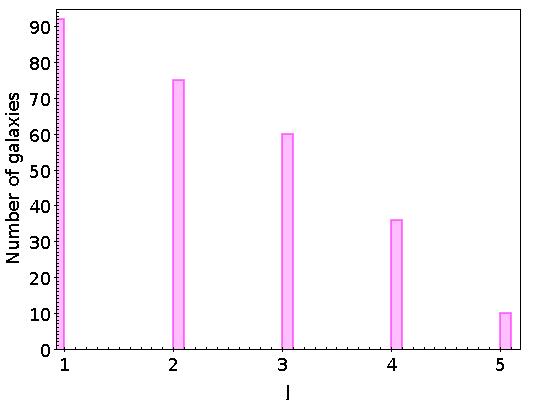}
\end{center}
\caption{Histogram of the J classification for the 273 RPS candidates of Sample 2.}
\label{fig:histoJ_zphot}
\end{figure}

\begin{figure}
\begin{center}
\includegraphics[width=0.3\textwidth]{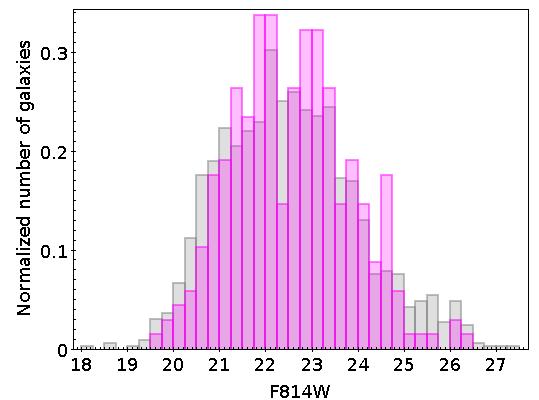}
\end{center}
\caption{Magnitude histograms for the 1326 galaxies in the [0.48,0.61] redshift interval in grey and for the 273 RPS candidates of Sample~2 in magenta, normalized to the same area.
}
\label{fig:histomag_zphot}
\end{figure}

The histograms of the redshifts, J values and F814W magnitudes are shown in Figs.~\ref{fig:histo_zphot}, \ref{fig:histoJ_zphot} and \ref{fig:histomag_zphot} for the 1326 galaxies in the [0.48,0.61] photometric redshift interval and for the 273 RPS candidates of Sample~2.

The median photometric redshifts of the two samples are 0.5394 and 0.5441 for all galaxies and RPS candidates of Sample 2 respectively. A KS test does not reject the hypothesis that these two samples  are drawn from the same parent population at a probability of 12\%.
The median F814W magnitudes of the two samples are 22.44 and 22.53 for all galaxies and RPS candidates of Sample 2 respectively. A KS test does not reject the hypothesis that these two samples  are drawn from the same parent population at a probability of 6.4\%.

One can note that there are more faint galaxies in Sample 2 than in Sample 1, since it is obviously easier to obtain photometric redshifts than spectroscopic redshifts.

\section{Spatial and phase space distributions}

\subsection{Sample 1 }

\subsubsection{Spatial distribution}

\begin{figure*}
\begin{center}
\includegraphics[width=0.8\textwidth]{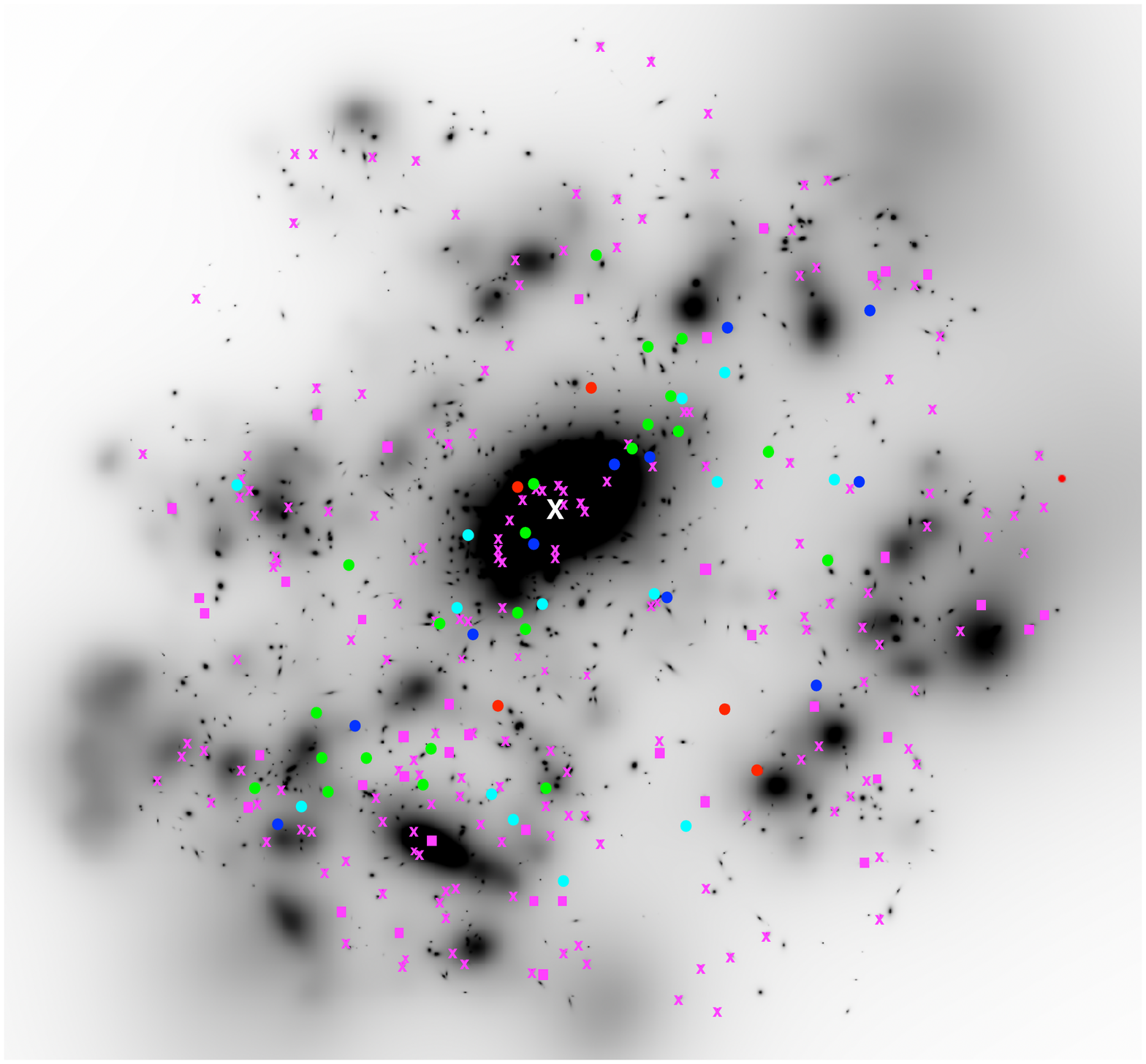} 
\end{center}
\caption{Figure~4 from \citet{Tam+20} showing the large scale-structure in a total region of $30\times 30$~arcmin$^2$ around MS0451. 
The positions of RPS candidates of Samples 1a, 1b, 1c, and 1d are respectively shown as green, cyan, blue and red circles. The RPS candidates of Sample 2 with grades J=1 to 3 are shown as magenta crosses, and those with J=4 and 5 with magenta squares. The white cross indicates the position of the cluster centre. North is up and east is left.}
\label{fig:Tamjelly2}
\end{figure*}

\begin{figure*}
\begin{center}
\includegraphics[width=0.8\textwidth]{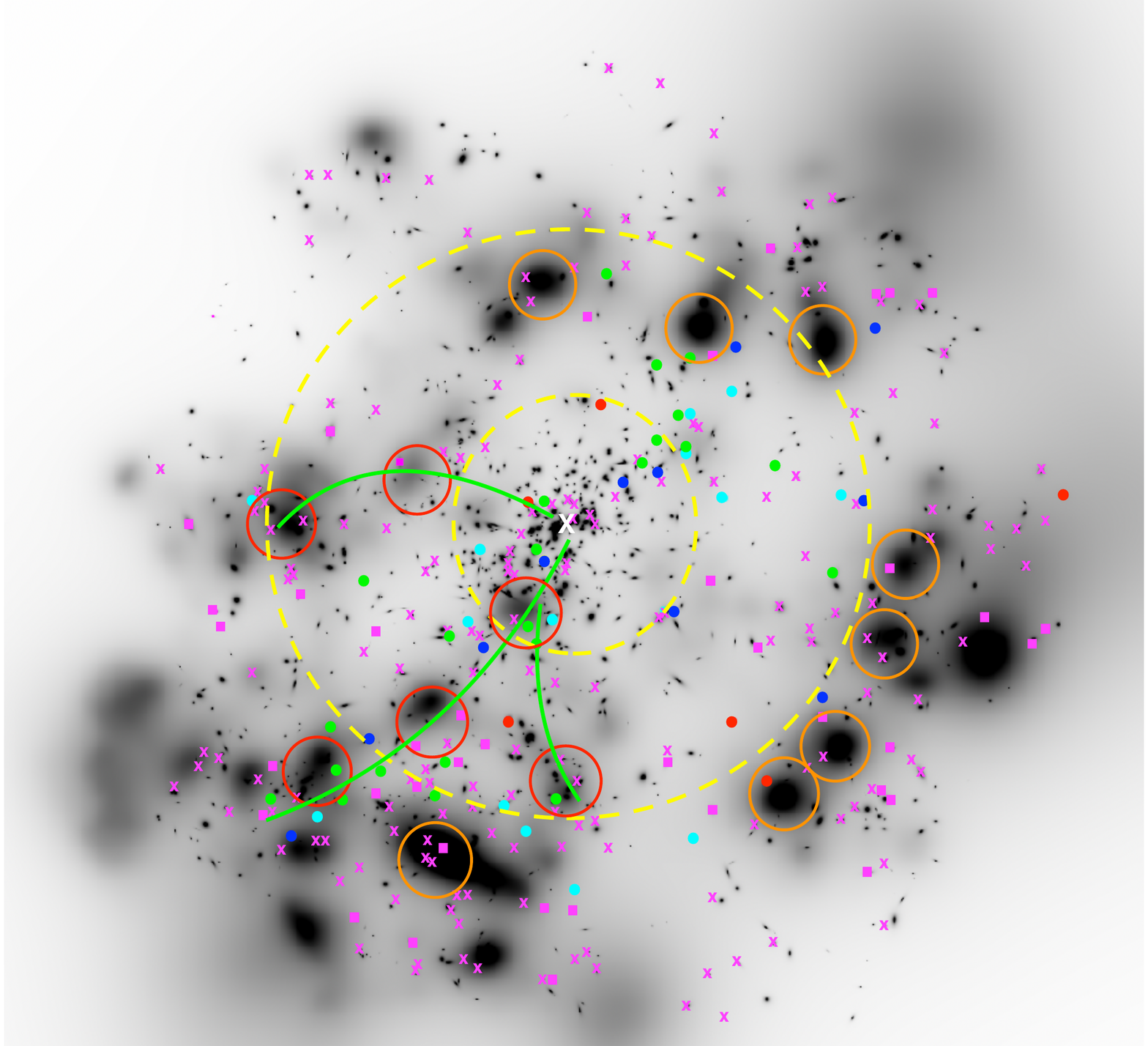} 
\end{center}
\caption{Figure~9 from \citet{Tam+20} showing the low density large scale structure in a total region of $30\times 30$~arcmin$^2$ around MS0451 with the substructures (red circles) and three filaments (in green) superimposed. The orange circles are unlikely to be cluster substructures, and are possibly the projection of large-scale structures at other redshifts along the line of sight. 
The positions of RPS candidates of Samples 1a, 1b, 1c and 1d are respectively shown as small green, cyan, blue and red circles. The RPS candidates of Sample 2 with grades J=1 to 3 are shown as magenta crosses, and those with J=4 and 5 with magenta squares. The white cross indicates the position of the cluster centre. The two yellow dashed circles have radii of $0.6\times r_{200}$ and $1.4\times r_{200}$, corresponding to the peaks seen in Fig.~\ref{fig:historsurr200}. North is up and east is left.}
\label{fig:Tamjelly}
\end{figure*}

\citet{Tam+20} combined strong and weak lensing constraints to reconstruct the density field of MS0451 in a $6 \times 6$~Mpc$^2$ area. For the cluster centre, they adopted the best-fit strong lensing mass model from \cite{Jauzac+21}. For the large-scale structures, they measured the weak lensing shear signal of background galaxies from the HST images and reconstructed the projected (2D) mass distribution by using LENSTOOL \citep{Jullo+07,JulloKneib09} and the method presented in \cite{Jauzac+15a}. The total mass distribution reconstructed from their joint strong and weak lensing analysis is shown in \cite{Tam+20} Figure~4, and the low density large scale structure with strong lensing potentials subtracted appears in their Figure~9. In the reconstructed large scale environment, there were 14 weak lensing peaks with S/N>3 detected. \cite{Tam+20} further investigated the validity of these overdense regions by considering the redshift information. They found that only 6 of these 14 weak lensing peaks were at the cluster redshift. These were then considered as substructures, constituting a part of the extended cluster (they are noted as red circles in Fig.~\ref{fig:Tamjelly}, while the others were considered as projections of structures at other redshifts along the line of sight (the orange circles in Fig.~\ref{fig:Tamjelly}). Based on the distribution of these six cluster substructures, \citet{Tam+20} proposed that three filaments (in green in Fig.~\ref{fig:Tamjelly}) are connected to the cluster core. 

The large scale structure drawn by \citet{Tam+20} is shown in Fig.~\ref{fig:Tamjelly2} with our four spectroscopic samples of RPS candidates (Samples 1a, 1b, 1c, and 1d) drawn in different colours, according to their J$_{mean}$ class (likeliness to be a RPS galaxy), as well as the photometric sample (Sample~2) in magenta, with different symbols for J$\le 3$ and for J$>3$ (we address the photometric sample in \ref{sec:S2}). 
We can see that overall the spectroscopic RPS candidates 
are spread over the entire mosaic, and
only a few of them are located in the dense central cluster region. This contrasts with the distribution of all spectroscopic cluster members that show the typical higher concentration toward the cluster core (RPS candidates being more sparsely distributed).
The RPS candidates show a somewhat higher concentration along the southeast to northwest cluster elongation axis, but in view of the small numbers involved it is difficult to make a statistical analysis of the significance of this observation.
If one now looks at the low density large scale structure drawn by \citet{Tam+20} displayed in Fig.~\ref{fig:Tamjelly}, we can see that RPS candidates tend to avoid the substructures highlighted as red circles, and they show no trend of being spatially linked to the three filaments. 
This is discussed further in Sect.~\ref{sec:discu}.

\begin{figure}
\begin{center}
\includegraphics[width=0.3\textwidth]{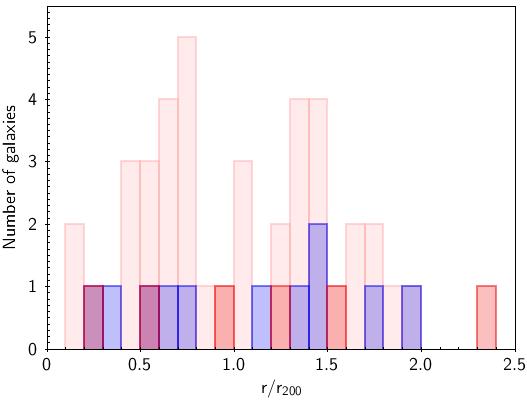}
\end{center}
\caption{Histogram of the distances of the 56 spectroscopic RPS candidates to the cluster centre in units of $r_{200}$. Samples~1a and 1b are shown in light pink, 1c is blue and 1d is red.}
\label{fig:historsurr200}
\end{figure}

The histogram of the distances of the 56 RPS Sample 1 candidates to the cluster centre in units of $r_{200}$ is displayed in Fig.~\ref{fig:historsurr200}, with $r_{200} = 2110$~kpc \citep{Durret+21}. We can see that there are only five RPS candidates at small clustercentric distances ($r<0.4\times r_{200}$). There are two peaks in the distribution around 0.6$\times r_{200}$ and 1.4$\times r_{200}$.
There is no obvious feature at a clustercentric radius of 0.6$\times r_{200}$. On the other hand, substructures shown as red circles in Figs.~\ref{fig:Tamjelly2} and \ref{fig:Tamjelly} fall close to a clustercentric radius of 1.4$\times r_{200}$, shown in  Fig.~\ref{fig:Tamjelly} by the outer dashed yellow circle. One may note that this is also the case for orange circles, though \citet{Tam+20} consider that they are probably not structures belonging to the cluster.
An excess of RPS candidates at this clustercentric radius may therefore be linked to an excess of cluster galaxies in that region (that we are unable to probe with our data).  

We note that Fig.~\ref{fig:historsurr200} is quite similar to the analogous figure for MACS0717 (Fig. 6 left, in \cite{Durret+21}, where an absence of RPS candidates in the cluster core was also noted. We further discuss this in our final remarks.

\subsubsection{Phase space distribution}
\label{sec:phase}

\begin{figure*}
\begin{center}
\includegraphics[width=0.4\textwidth]{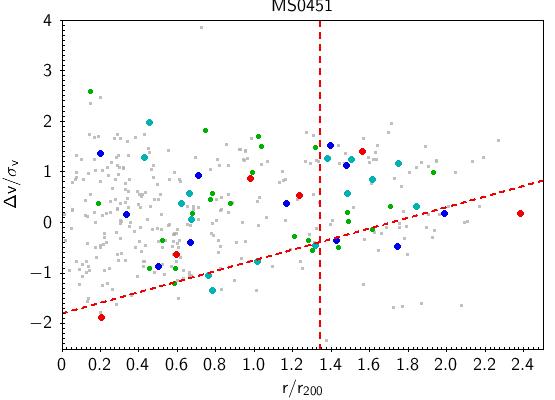}  \kern0.1cm%
\includegraphics[width=0.4\textwidth]{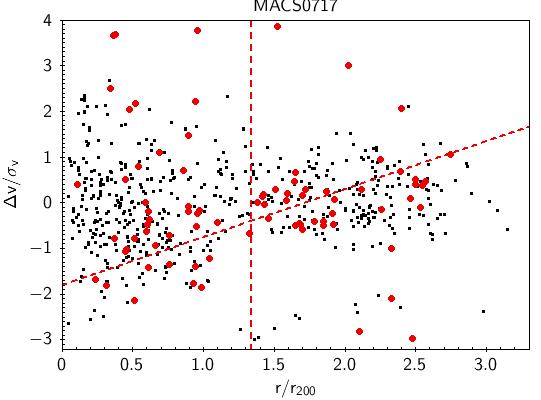}
\end{center}
\caption{Projected phase space diagrams (see text) for MS0451 (left) and MACS0717 (right). 
For MS0451, the 359 galaxies with spectroscopic redshifts in the cluster range are in grey, and  RPS candidates of samples 1a, 1b, 1c and 1d are respectively shown as small green, large green, blue, and red points. 
For MACS0717, all galaxies in the cluster range are in black and RPS candidates are in red. The vertical and oblique dashed lines are those of figure 9 of \cite{Mahajan+11} as explained in the text.
}
\label{fig:phasespace}
\end{figure*}

\begin{table*}
\centering
\caption{Percentages of RPS candidates in the virialized (limit ar $r=r_{100}$), infall and backsplash regions of MS0451 and MACS0717.}
\begin{tabular}{l|rrrr|rrrr}
\hline \hline
Region &     & ~~~MS0451 &         & &     & ~~~~MACS0717 & & \\
       & RPS &             & non-RPS & & RPS &              & non-RPS & \\
\hline
virialized & 37/56 & 66\%~~~~~~~~~~~~ & 282/359 & 79\% & 40/81 & 49\%~~~~~~~~~~~~ & 347/557 & 62\% \\
infall     &  6/56 & 11\%~~~~~~~~~~~~ & 32/359 &  9\%  & 25/81 & 31\%~~~~~~~~~~~~ & 136/557 & 24\%  \\
backsplash & 13/56 & 23\%~~~~~~~~~~~~ & 45/359 & 12\%  & 16/81 & 20\%~~~~~~~~~~~~ &  74/557 & 13\% \\
\hline
\end{tabular}
\label{tab:Mah}
\end{table*}

In order to get more insight on the dynamical properties of the RPS candidates, we drew projected phase space diagrams for  MS0451 and for MACS0717, as a comparison, and since both clusters lie at z$\sim$0.5.

We chose MACS0717 as a comparison cluster because it is a well-studied very massive cluster \citep{Ebeling+07} at a redshift (z=0.54) similar to that of MS0451, with deep and extended HST observations, since it is part of the Hubble Frontier Fields survey \citep{Lotz+17}. It is a very strong gravitational lens, and a lensing analysis coupled with X-ray data has revealed a complex merging system involving four cluster-scale components \citep{Ma+09,Zitrin+09,Limousin+12,Limousin+16}. A long filament extending southeast of the cluster core has been detected in the 3D distribution of galaxies \citep{Ma+08}, of red-sequence galaxies \citep{Durret+16}, and through weak lensing \citep{Jauzac+12,Medezinski+13,Martinet+16,Jauzac+18}. Based on HST data of comparable depth as those presented here (but for MACS0717 we had images in the two filters F555W and F814W), a search for RPS galaxies in MACS0717 has been presented by \citet{Durret+21}, one of the two galaxy classifiers being the same in both studies (FD).

To draw the phase diagrams of these two clusters, we computed the values of $\Delta v/\sigma _v$, where $\Delta v$ is the difference between each galaxy velocity and the mean cluster velocity, and drew these values as a function of the projected distance to the cluster centre in units of $r_{100}$, $r/r_{100}$. For MS0451, we have previously estimated $r_{200}=2110$~kpc \citep{Durret+21}, and from this value we estimate $r_{100}=(1.34\pm 0.04)\times r_{200} = 2827$~kpc (G. Mamon, private communication). 
Based on a weak lensing analysis, \citet{Tam+20} give a mass  M$_{200}=1.65\times 10^{15}$~M$_\odot$ for MS0451.
The cluster velocity dispersion can then be computed using equation (1) from \cite{Munari+13}:
$$\sigma_v  = 1090 \times [h(z) \times M_{200}]^{(1/3)},$$
\noindent
where $h(z) = H(z) / 100$, $M_{200}$ is expressed in units of $10^{15}$~M$_\odot$, and $\sigma_v$ is the unidimensional velocity dispersion in units of km/s, and we obtain $\sigma_v=1205$~km~s$^{-1}$, as in \citet{Durret+21}.
The result is shown in Fig.~\ref{fig:phasespace}, together with a similar figure for MACS0717, for which the corresponding values are given by \citet{Durret+21}. 

The plots in Fig.~\ref{fig:phasespace} can be compared to results by \citet{Mahajan+11}, who define classes based on the radial phase-space distribution of particles. For them, particles that are outside the virial radius ($r_{100}$), and travelling at velocities lower than the critical velocity, are falling into the cluster for the first time (infalling), while particles outside the virial radius with higher velocities must have crossed the cluster at least once (backsplash).
The critical velocity is the radial velocity that separates infalling galaxies entering the cluster for the first time from galaxies entering or crossing the cluster for the second time (or more). It is shown as an oblique red line in Fig.~9 of \citet{Mahajan+11} and is defined by their equation (2). 

In their Fig.~9,  \citet{Mahajan+11} separate virialized, infall, and backsplash galaxies with cuts taken from \citet{Sanchis+04} in units of $r_{100}$. In a similar way, we draw the limits between the virialized and non-virialized regions as lines in Fig.~\ref{fig:phasespace} for the two clusters.
The oblique lines separating infall and backsplash galaxies were drawn as in \citet{Mahajan+11} Figure~9. 

For MS0451, applying the \citet{Sanchis+04} cuts to our data, the numbers and percentages of RPS candidates are given in Table~\ref{tab:Mah}:
66\% are located in the virialized region ($r/r_{200}\leq 1$), and 34\% have $r/r_{200}>1$: 11\% of the total sample in the infall region and 23\% in the backsplash region.
The corresponding numbers for non-RPS galaxies are: 79\% in the virialized region, and for the remaining 91 galaxies, 9\% of the total sample in the infall region and 12\% in the backsplash region.

For MACS0717, 49\% of the RPS candidates are in the virialized region, 31\% in the infall region and 20\% in the backplash region. Fpr non-RPS galaxies, 62\% are in the virialized region, 24\% and 13\% in the infall and backsplash regions. These percentages will be discussed in Section~\ref{sec:discu}.

\subsection{Sample 2 }\label{sec:S2}

The  spatial distribution of RPS candidates in Sample~2 is shown in Figs.~\ref{fig:Tamjelly2} and \ref{fig:Tamjelly}. 
They are found over the entire cluster field of view but those with J$\le 3$ in particular do not tend to avoid the dense zones as much as RPS candidates of Sample~1, possibly because some of these objects are not cluster members, and are just seen in projection on the cluster. These are also the candidates with less obvious RPS features.

Alternatively, and since these are generally less luminous, thus less massive galaxies when compared to Sample 1 (as confirmed in Section 4), this could highlight a real difference in the spatial distribution of different mass RPS candidates.

\section{Stellar populations of RPS candidates}

\subsection{Method}

We used LePhare \citep{Ilbert+06}, through the GAZPAR interface\footnote{https://gazpar.lam.fr/home}, to fit the spectral energy distributions (SEDs) of the four samples. Sample~1 includes the 56 RPS candidates with spectroscopic redshifts. Sample~2 includes the 273 RPS candidates with photometric redshifts in the cluster range. Sample~3 includes 282 non-RPS galaxies with spectroscopic redshifts in the same range as Sample~1. Sample~4 includes 1027 non-RPS galaxies with photometric redshifts in the same range as Sample~2.
The fits were made with the \citet{BC03} models assuming a \citet{Chabrier+03} initial mass function. Based on an input catalogue with positions, eight magnitudes and corresponding errors, and redshifts, LePhare fits the galaxy SEDs and infers from the best fit template for each galaxy the stellar mass, mean stellar population age, star formation rate (SFR), and specific star formation rate (sSFR). 
The input parameter space was carefully selected so as to cover the expected characteristics of late-type galaxies with probable star forming activity as well as the SEDs of the typically abundant early-type galaxies in clusters.

Out of the 56 spectroscopically confirmed RPS candidates, the best fit template spectrum includes all the main emission lines in the optical ([OII]3727, [OIII]4959, 5007, H$\beta$ and H$\alpha$) for 33 of them, and
an H$\alpha$ (and weak H$\beta$) line for 20 of them (see examples of fits in Fig.~7 of \citet{Durret+21}). Only three of the best fit template spectra include no emission line. This implies that the majority of our RPS candidates are forming stars.

The histograms of the stellar masses, ages, SFRs and sSFRs are respectively shown in Figs.~\ref{fig:mass}, \ref{fig:age}, \ref{fig:SFR}, and \ref{fig:sSFR} for Samples 1, 2, 3, and 4, and will be discussed in detail along the next sub-sections.

\subsection{Stellar masses}
\label{subsec:masses}

\begin{figure}
\begin{center}
\includegraphics[width=0.4\textwidth]{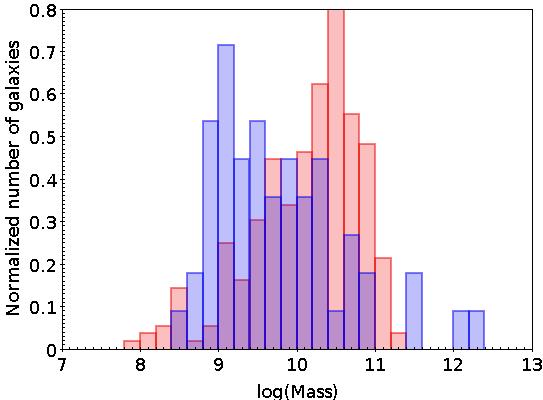}  
\includegraphics[width=0.4\textwidth]{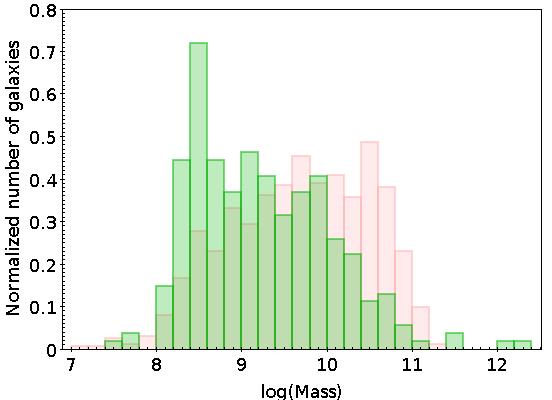} 
\end{center}
\caption{Histograms of the stellar masses (in units of solar masses) normalized to the same area. Top: Samples 1 and 3 (spectroscopic redshifts), with RPS candidates in blue and non-RPS galaxies in red. Bottom: Samples 2 and 4 (photometric redshifts), with RPS candidates in green and non-RPS galaxies in light pink. }
\label{fig:mass}
\end{figure}

Figure~\ref{fig:mass} shows that RPS candidates (Samples 1 and 2) appear to have lower mass galaxies than non-RPS galaxies (Samples 3 and 4).
The median masses for Samples 1 and 2 are respectively $10^{9.57}$ and $10^{9.15}$~M$_\odot$, while they are $10^{10.28}$ and $10^{9.73}$~M$_\odot$ for Samples 3 and 4. Differences between mass distributions of RPS and non-RPS candidates are confirmed statistically by a KS test, the results of which are given in Table~\ref{tab:KS} (as for all other quantities).

If we now compare the masses of Sample~1 and Sample~2 galaxies, a KS test also confirms that these two distributions differ.
The fact that galaxies of Sample 1 are predominantly more massive than those of Sample 2 is not surprising since galaxies observed spectroscopically are intrinsically brighter, and thus likely to be more massive. As expected, our photometric redshift samples allow us to probe the low mass regime, missed in the bright spectroscopic samples.

In order to understand if completeness issues are affecting these comparisons, we apply a cut in mass, limiting the samples to 
$10^9\leq$M$\leq 10^{10.5}$~M$_\odot$ (frontier values suggested by Fig.~\ref{fig:mass} - top).
The median masses for Samples 1 and 2 are then respectively $10^{9.57}$ and $10^{9.63}$~M$_\odot$, while they are $10^{9.97}$ and $10^{9.79}$~M$_\odot$ for Samples 3 and 4. KS tests confirm that all these samples are likely drawn from different populations, so the above inferences should hold, and RPS galaxies appear to have, in general, lower masses than non-RPS galaxies.
Although uncertainties in the mass determination are large, 
the trend we find agrees with what is observed in the local Universe across groups and clusters, as probed at z<0.04 in the SDSS-UNIONS survey by \cite{Roberts+22}: RPS candidates commonly seem to be low-mass galaxies.

At similar redshifts,
the mass distribution of the 81 spectroscopic RPS candidates in MACS0717 (Fig.~8 in \cite{Durret+21}) was found to be in
the $10^9\leq$M$\leq 10^{11}$~M$_\odot$ range, while in MS0451, they are lower mass galaxies, with 16\% of Sample~1 galaxies having masses below $10^9$~M$_\odot$. 
HST imaging of MS0451 is deeper than for MACS0717. This difference in depth could explain the difference in RPS mass ranges observed between MS0451 and MACS0717.

\cite{Moretti+22} observed RPS galaxies in the merging clusters A2744 and A370, at $z\sim 0.3-0.4$ with VLT/MUSE and found masses in the $10^8\leq$M$\leq 10^{10.6}$~M$_\odot$ range, 
but comparing our Sample 1 values with theirs is hazardous due to a different selection procedure of RPS galaxies (indeed, their project specifically identified RPS galaxies with tails), different observational limits (e.g. in magnitude and cluster radial coverage), and different SED fitting codes to derive stellar masses. We thus refrain from drawing any conclusion.

\subsection{Stellar ages}
\label{subsec:ages}

\begin{figure}
\begin{center}
\includegraphics[width=0.4\textwidth]{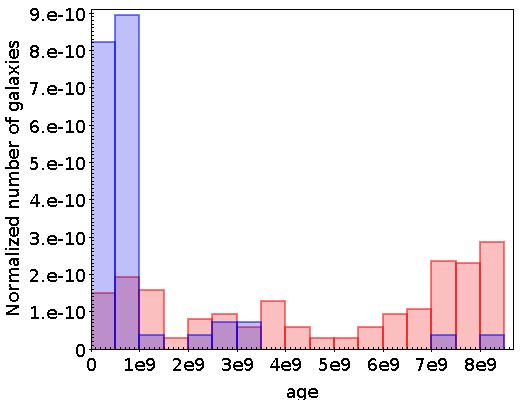}  \kern0.1cm%
\includegraphics[width=0.4\textwidth]{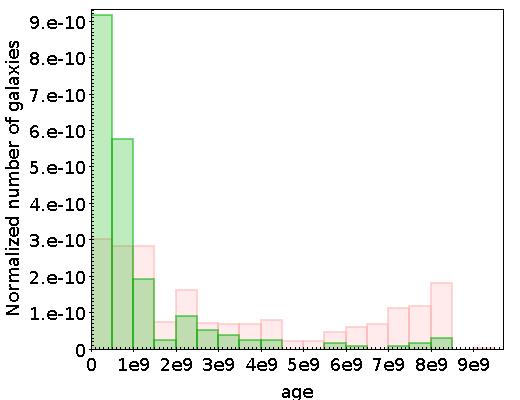} 
\end{center}
\caption{Same as Fig.~\ref{fig:mass}, but for stellar ages (in years).}
\label{fig:age}
\end{figure}

To estimate the accuracy of the stellar ages estimated by Le Phare, we considered for the galaxies of Samples 1 and 3 the quantity $(age_{max}-age_{min})/2$, where 
$age_{max}$ and $age_{min}$ are the upper and lower limits on the age given by Le Phare. We find that this quantity is much smaller for RPS than for non-RPS galaxies. For RPS galaxies, it is smaller than $2.5\times 10^8$~yrs for 30\% of the sample and smaller than $5\times 10^8$~yrs for 79\% of the sample. On the other hand, for non-RPS galaxies the error on the stellar age is smaller than $5\times 10^8$~yrs for only 8\% of the sample. The percentages of non-RPS galaxies with age errors smaller than $10^9$~yrs, $1.5\times 10^9$~yrs, and $2\times 10^9$~yrs, are respectively 48\%, 70\%, and 87\%.

The mean stellar age distributions (Fig.~\ref{fig:age}) show that RPS candidates (Samples 1 and 2) are predominantly younger than non-RPS galaxies (Samples 3 and 4). KS tests show that the differences between the distributions are statistically significant, so we can infer that RPS candidates typically have less evolved stellar populations than non-RPS galaxies.  
The median ages for Samples 1 and 3 are $5.4\times 10^8$ and $5.5\times 10^9$~yrs respectively, while the corresponding values for Samples 2 and 4 are $5.1\times 10^8$ and $2.3\times 10^9$~yrs. 

However, the difference in age distribution between RPS candidates and non-RPS galaxies could simply be driven by the different mass distributions of these galaxies (discussed in the previous sub-section). We thus redid our analysis after limiting the samples to 
$10^9\leq$M$\leq 10^{10.5}$~M$_\odot$ (as before). We then find  that
the median ages for Samples 1 and~3 are $5.1\times 10^8$ and $3.0\times 10^9$~yrs respectively, while the corresponding values for Samples 2 and 4 are $8.1\times 10^8$ and $2.3\times 10^9$~yrs. As found without applying a mass cut, KS tests still imply that RPS candidates are typically younger than non-RPS candidates.

If we compare the age distribution of Sample~1 to that of RPS candidates in MACS0717, we find that RPS candidates tend to be younger in MS0451: 86\%  of the galaxies in Sample 1 have mean ages younger than $10^9$~yrs, while this percentage is only 29\% in MACS0717. A cut in mass 
($10^9\leq$M$\leq 10^{10.5}$~M$_\odot$) respectively changes these percentages to 64\% and 27\%.

\subsection{Star formation rates}
\label{subsec:SFR}

\begin{figure}
\begin{center}
\includegraphics[width=0.4\textwidth]{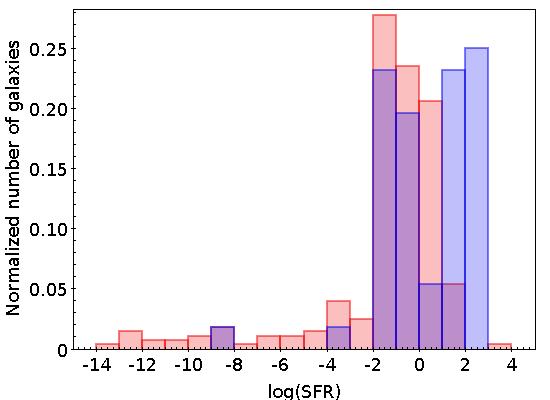}  \kern0.1cm%
\includegraphics[width=0.4\textwidth]{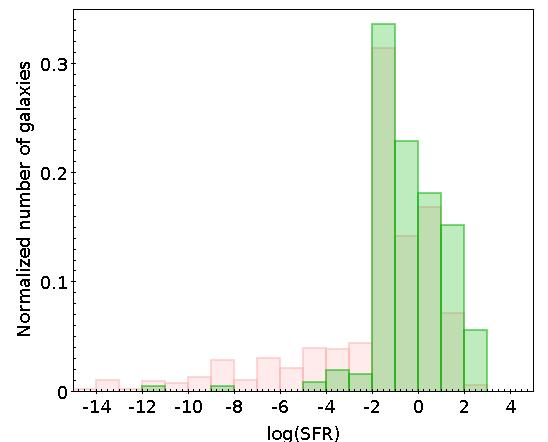} 
\end{center}
\caption{Same as Fig.~\ref{fig:mass} for star formation rates (in units of M$_\odot$~yr$^{-1}$). A few non-RPS galaxies with even lower SFR are not shown to avoid shrinking too much the histograms.}
\label{fig:SFR}
\end{figure}

The SFR distributions of RPS and non-RPS galaxies look different (Fig.~\ref{fig:SFR}), as  there is a low SFR tail in non-RPS galaxies (both Samples 3 and 4) that is almost absent when considering RPS candidates (Samples 1 and 2).

This is also hinted by the median values of the distributions, which are $10^{0.679}$ and $10^{-1.004}$~M$_\odot$~yr$^{-1}$ for Samples 1 and 3 respectively, and $10^{-0.75}$ and $10^{-1.26}$~M$_\odot$~yr$^{-1}$ 
for Samples 2 and 4.
Indeed, KS tests confirm that SFR distributions of Samples 1 and 3 on the one hand, and of Samples 2 and 4 on the other hand, differ. 
Therefore, the global trend for higher SFR in RPS candidates compared to non-RPS galaxies is statistically significant. 

As before, we check the effect of a mass cut 
($10^9\leq$M$\leq 10^{10.5}$~M$_\odot$)
on the SFR distributions. We then find median values of $10^{-0.05}$ and $10^{-0.965}$~M$_\odot$~yr$^{-1}$ for Samples 1 and 3 respectively, and $10^{0.718}$ and $10^{-1.33}$~M$_\odot$~yr$^{-1}$ 
for Samples 2 and~4. KS tests again confirm that these samples are not drawn from the same population. 

The SFRs measured for RPS galaxies in MACS0717 show values between $10^{-1.0}$ and $10^{1.8}$~M$_\odot$~yr$^{-1}$, with a median of $10^{0.625}$~M$_\odot$~yr$^{-1}$, comparable to that of MS0451.
However, if one compares SFR histograms of Fig.~\ref{fig:SFR} for Sample~1 of MS0451 and Fig.~9 (left) in \cite{Durret+21} for MACS0717, one can note that MS0451 has 22 galaxies (out of 56) with a SFR larger than the higher value observed in MACS0717 (SFR=$10^{1.75}$~M$_\odot$~yr$^{-1}$).

If we apply a mass cut  
($10^9\leq$M$\leq 10^{10.5}$~M$_\odot$)
to MACS0717, we find a median SFR of $10^{0.73}$~M$_\odot$~yr$^{-1}$, much higher than the corresponding value of $10^{-0.708}$~M$_\odot$~yr$^{-1}$ estimated for Sample 1 with the same cut. However, in view of the broader distribution of masses in MS0451, a comparison of the sSFRs of the two clusters is probably more relevant (see below).

In the \cite{Moretti+22} study of the galaxy populations of clusters A2744 and A370, dynamically similar to ours but at lower redshifts, they measure SFR values up to  73~M$_\odot$~yr$^{-1}$ (i.e. $\sim 10^{1.86}$~M$_\odot$~yr$^{-1}$) for their RPS candidates. Figure~\ref{fig:SFR} reveals that 18 RPS candidates of Sample 1 
(i.e. 32\%) and 19 RPS candidates of Sample 2 (7\%) exceed this value - an effect that, uncertainties apart, can be simply due to the redshift difference and to the likely larger gas reservoirs of galaxies at z $\sim0.5$.

It would be very interesting to do a spatially resolved assessment of the star formation within the galaxies but, unfortunately, the lack of spatially resolved data (spectroscopy and/or magnitudes) prevents us from estimating the SFR in galaxy disks and tails separately, as done e.g. in \cite{Gullieuszik+20} based on VLT/MUSE data.

\subsection{Specific star formation rates}
\label{subsec:sSFR}

\begin{figure}
\begin{center}
\includegraphics[width=0.4\textwidth]{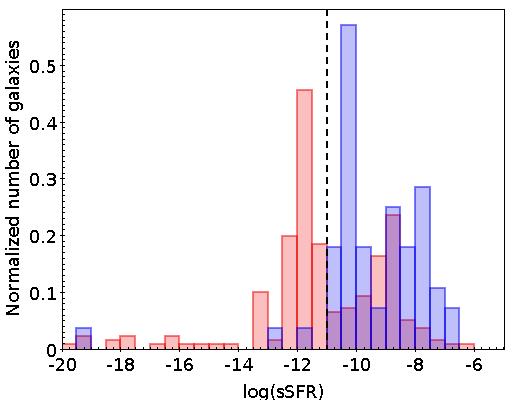}  \kern0.1cm%
\includegraphics[width=0.4\textwidth]{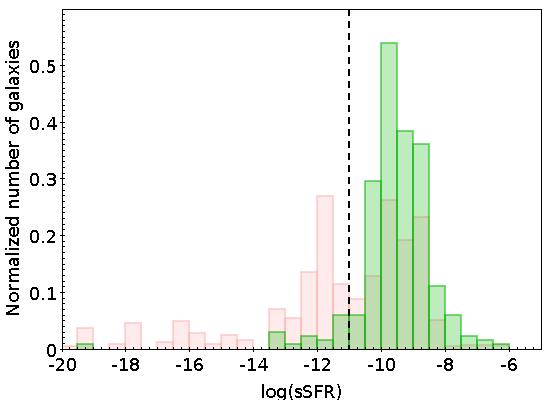} 
\end{center}
\caption{Same as Fig.~\ref{fig:mass} for specific star formation rates (in units of yr$^{-1}$). The vertical dashed line signals the commonly used limit between quiescent and star-forming galaxies.}
\label{fig:sSFR}
\end{figure}

Specific star formation rates should be more informative than plain SFR since they are normalized to galaxy stellar masses. Figure~\ref{fig:sSFR} shows that
RPS candidates appear to typically have larger sSFR than non-RPS galaxies. Also, only 5\% and 7\% of galaxies in Samples 1 and 2 have log(sSFR)$<-11$, the usual limit that separates non star forming galaxies from star forming ones. On the other hand, 63\% and 51\% of galaxies in Samples 3 and 4 have log(sSFR)$<-11$ (these likely make up the quiescent galaxy population of the cluster). Noticeably, Sample 1 shows a somewhat bimodal sSFR distribution, 
with a non-negligible fraction of high sSFR galaxy members, which are commonly absent in lower-z clusters especially in their cores \citep[e.g.][]{Oemler74, Dressler80, Goto+03, Kauffmann+04, Blanton+05}.

The differences between the typical sSFRs of RPS and non-RPS galaxies is confirmed by the median values of the distributions, $10^{-9.57}$ and $10^{-11.55}$~yr$^{-1}$ for Samples 1 and 3 respectively, and $10^{-9.57}$ and $10^{-11.12}$~yr$^{-1}$ for Samples 2 and 4, respectively, and the differences between the samples are confirmed by KS tests.

When we apply a mass cut as before 
($10^9\leq$M$\leq 10^{10.5}$~M$_\odot$), we find median values of sSFR of $10^{-9.32}$ and $10^{-11.10}$~yr$^{-1}$ for Samples 1 and 3 respectively, and $10^{-8.92}$ and $10^{-11.33}$~yr$^{-1}$ for Samples 2 and 4 respectively, and the differences between the samples are here again confirmed by KS tests.
Therefore, the higher typical sSFRs of RPS candidates compared to non-RPS galaxies seem to be statistically significant.

In MACS0717, the median sSFR of RPS candidates is $10^{-9.36}$~yr$^{-1}$, and 10\% of these galaxies have log(sSFR)$<-11$, results
which are perfectly similar to those derived for Sample~1 of MS0451. If we apply a mass cut 
($10^9\leq$M$\leq 10^{10.5}$~M$_\odot$), the median sSFR of MACS0717 becomes $10^{-9.13}$~yr$^{-1}$, 
probably not significantly different from the value of $10^{-9.79}$ for MS0451, within uncertainties. The sSFR histograms of MS0451 and MACS0717 also look similar. 

To summarize these results, RPS candidates and non-RPS galaxies do not seem to be
drawn from the same parent population, as confirmed by KS tests. As already mentioned by  \citet{Durret+21} and references therein, RPS candidates tend to have smaller masses, younger mean stellar ages, and larger SFRs and sSFRs than non-RPS galaxies. This agrees with the general picture of these galaxies having star formation triggered by their interaction with the hot intracluster gas, and our findings for MS0451 are consistent with this hypothesis.

\subsection{Star formation rate to mass relation}
\label{subsec:ML}

\begin{figure}
\begin{center}
\includegraphics[width=0.48\textwidth]{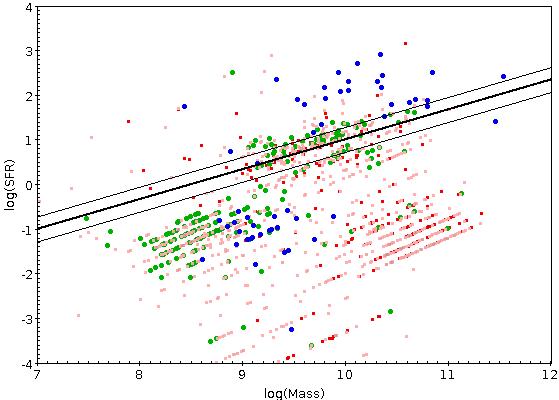} 
\end{center}
\caption{SFR as a function of stellar mass for samples 1 (blue), 2 (green), 3 (red), and 4 (light pink). The black line indicates the relation by \citet{Vulcani+16} and its approximate dispersion of $\pm 0.28$ (light black lines).
}
\label{fig:mass_SFR}
\end{figure}

A plot of the SFR as a function of stellar mass is shown in Fig.~\ref{fig:mass_SFR} for our four samples, together with the main sequence (MS) of star forming galaxies established by \citet{Vulcani+16} for a sample of 76 galaxies belonging to 10 clusters in the $0.3<z<0.7$ redshift range, and spanning a similar stellar mass range as ours, $10^{8.25} \la $~M$ ~\la 10^{11}$~M$_\odot$. 
We note that the striped pattern that appears in Fig.~\ref{fig:mass_SFR} is due to the limited number (12) of templates used by LePhare \citep{Ilbert+06}, namely in what concerns star formation histories and ages. This is especially the case for typical SEDs of low or non star forming galaxies, for which there are only two templates (that, moreover, do not consider dust attenuation).
The ensemble of templates should, however, cover all the parameter space conveniently while avoiding degeneracies among too similar models (that might hamper convergence and would not provide better results for our data, given the usual uncertainties associated with estimating SFRs from template fitting).

The percentages of galaxies - separated by samples and having M$\leq 10^{10.5}$~M$_\odot$ - located above, within, and below the Vulcani MS are provided in Table~\ref{tab:vulcani}. We have added the corresponding percentages for the RPS candidates found in MACS0717 (these  should be directly comparable to Sample 1 of MS0451). These numbers are meant to guide the eye but one should however consider that they can only indicate possible trends, given the usual uncertainties associated with the estimates of both stellar mass and star formation rates. With this in mind, we can still infer that, globally, RPS candidates seem to be forming stars more actively than their non-RPS counterparts (49\% and 43\% of Samples 1 and 2 lie within or above the star forming main sequence, against just 37\% and 29\% for Samples 3 and 4). This trend was more strongly marked for MACS0717, where only 21\% of the RPS candidates lie below the MS. The other difference between the two clusters seems to be the inversion of RPS candidates within and above the MS: MACS0717 has both regions equally populated whereas, if we take Sample 1 of MS0451, only 4\% of the 49\% previously mentioned are found between the thin black lines in Fig.~\ref{fig:mass_SFR} that delimit the region populated by galaxies with normal star formation rates for their stellar masses.

\begin{table*}[h!]
\begin{center}
\caption{Percentages of galaxies located above (top line) and within (bottom line) the \cite{Vulcani+16} main sequence (MS) for our four samples and for the RPS candidates of MACS0717, all the samples being cut at $10^9\leq$M$\leq 10^{10.5}$~M$_\odot$. }
\begin{tabular}{ c c c c c c c}
\hline
\hline
  & Sample 1 & Sample 2 & Sample 3 & Sample 4 & MACS0717 \\ 
\hline
above MS & 50\% & 47\% & 13\% & 18\% & 38\% \\  
in MS & 5\% & 32\% & 20\% & 20\% & 41\% \\  
\hline
\end{tabular}
\label{tab:vulcani}
\end{center}
\end{table*}

If we check the literature, we find that \cite{Cooper+22} derived the star-forming main sequence for 163 galaxies
in four EDisCS clusters in the redshift range $0.4 < z < 0.7$, out till the respective infall regions. Their points (see their Fig. 8), essentially in the mass range $[10^{8.5},10^{11.5}]$ M$_\odot$, are almost all below the Vulcani relation, confirming that a significant number of our RPS candidates typically have large SFRs for their stellar masses. 
In that same figure, we can see that the SFR values estimated by \cite{Finn+05} for galaxies belonging to three EDISCS clusters at $z\sim 0.75$ are above the Vulcani relation, particularly at low mass, so as mentioned by \cite{Finn+05} there is probably a redshift dependence, though there are other factors (such as mass) influencing the SFR.

If we now compare with low redshift observations targetting specifically RPS galaxies as well, our findings seem to agree with those of various authors. 
For example, \cite{Poggianti+16} detected 344 RPS candidates in 71 clusters (OMEGAWINGS+WINGS surveys comprising galaxies with stellar masses above $10^8$ M$_\odot$) and showed that they had increased SFRs (see their Fig.~8). 
\cite{Vulcani+18} found a similar result when comparing 42 RPS cluster galaxies to a control sample of 32 galaxies (field and undisturbed galaxies) from the GASP survey, all with stellar masses above $10^9$ M$_\odot$.
\citet{Roberts+22} observed a large sample of isolated field galaxies at low redshift and found that 77\% of their RPS galaxies (stellar masses above $10^{8.8}$ M$_\odot$) lie above the star-forming main sequence drawn in their Figure~8. The general assessment thus seems to be that environment,  particular RPS, is indeed triggering episodes of enhanced SFR at various redshifts.

Apart from these global trends, a closer inspection of Fig.~\ref{fig:mass_SFR} appears to indicate that there is a transition of behaviour at M$\sim 10^{9.4}$~M$_\odot$ in MS0451: more massive RPS candidates lie notably above the MS (having boosted SFRs), whereas lower mass RPS candidates actually populate preferably the lower part of the plane (appearing to have lost the capacity to form stars). This insight is confirmed by the median values of SFR: $10^{-0.94}$ and $10^{1.87}$~M$_\odot$~yr$^{-1}$ for M$\leq 10^{9.4}$~M$_\odot$ and M$> 10^{9.4}$~M$_\odot$ in Sample 1, $10^{-1.05}$ and $10^{0.97}$ in Sample 2, respectively. This dichotomy, especially marked in Sample 1, was already apparent in Fig.~\ref{fig:SFR} and could be an artefact of the SED fitting tool, a completeness issue, or a real trend. In the last case, a tentative  interpretation is that normal star forming galaxies with available gas reservoirs and large masses have boosted SFRs when subject to RPS (thus moving upward from the main sequence). In the lower mass regime, the increased SF episode leaves its marks on the galaxy (allowing its selection as having been affected by RPS), but either its gas is totally consumed in that episode or, due to the weaker gravitational potentials of these galaxies, what is left of it is rapidly swept away. Either way, subsequent SF is prevented.

This different behaviour for the different mass regimes of RPS candidates was not detected in low-z clusters \citep[e.g.][]{Poggianti+16} nor in the field \citep{Roberts+22} and will definitely need to be confirmed - or not - with a deeper and more complete set of spectroscopic data. If true, it might be linked to the specific dynamical conditions of this post-merging cluster. We note that the directly comparable study of MACS0717 was performed on a somewhat shallower dataset (see section \ref{subsec:masses}), which did not allow us to probe down to the same lower mass regime. This would have been especially interesting since MACS0717 is also a dynamically complex cluster - a merging system.
 
On a final note, we have looked at the distribution of the different J class galaxies in the SFR-mass plane: a simple separation between J$>3$ and J$\le3$ does not show any particular trend. The former, showing stronger RPS characteristics, are not distinguishable from the others in what regards their SFR-mass properties. We have also checked 
the positions of the 6 infall and 10 backsplash RPS candidates of Sample~1 in Fig.~\ref{fig:mass_SFR}: here, again, the different categories of galaxies are spread throughout the graph, without revealing any particular trend. These tentative analyses are, however, based on small numbers in each category, which renders any inference difficult.

\section{Discussion and conclusions} 
\label{sec:discu}

We present in this paper a search for RPS candidates in the post-merging cluster MS0451 located at redshift $z=0.5377$. This study is based on HST images covering a very large region reaching almost $2\times r_{200}$ (i.e. a spatial coverage reaching the cluster infall region), complemented with a large spectroscopic redshift catalogue and a magnitude catalogue in eight bands that allowed us to derive photometric redshifts. We compare our results to those obtained by \cite{Durret+21} in MACS0717, a massive merging cluster at similar redshift with comparable data, analyzed in the same way. Though spatially resolved spectroscopic data
would obviously give much more information, it is very challenging to acquire and analyze integral field datacubes for 
so many galaxies spread out till large clustercentric distances.
For example, \cite{Moretti+22} study two clusters at $z\sim 0.3-0.4$ with MUSE up to $(0.1-0.15)\times r_{200}$.

For MS0451, we find a percentage of RPS candidates with redshifts secured by spectroscopy of 56/359=15.6\%, which is close to the value of 81/557=14.5\% found for MACS0717 by \citet{Durret+21}. \citet{Tam+20} found that MS0451 is in a post-merger state, so it is not a relaxed cluster. However, MS0451 is less perturbed and extended than MACS0717. The fact that the percentages of RPS candidates in these two clusters are comparable could suggest that, at least for these two particular systems, the percentage of RPS galaxies is not correlated to the merging activity of the cluster, but there are some differences in their star formation properties - we shall come back to this issue below.
We can note that these percentages of RPS galaxies are within the range found by \citet{Roberts+22} in a sample of more than 50 low redshift clusters, based on data from the UNIONS survey. 

We show in Figs.~\ref{fig:Tamjelly2} and \ref{fig:Tamjelly} the overall and low density large scale distributions derived by \citet{Tam+20}, with the positions of our 56 spectroscopic RPS candidates (Sample~1) superimposed. We clearly see that there are few RPS candidates in the central dense cluster region, and that they completely avoid the Substructures numbered 1 to 14 in the \citet{Tam+20} paper. They do not particularly follow the filaments, since few of them are very close to the three filaments. 
For the 273 RPS galaxies of Sample~2 (photometric redshift sample) also shown in Figs.~\ref{fig:Tamjelly2} and \ref{fig:Tamjelly}, the spatial distributions of those with J classes of 4 and 5
also tend to avoid the dense regions, while this is not the case for galaxies with J$<3$. 
However, we must keep in mind that there must be some contaminants in Sample 2, that is to say foreground or background objects. If these are RPS galaxies, they will lead to an  overestimation of the number of RPS cluster members in Sample 2. On the other hand, if their distorted morphologies are due to other processes that are especially relevant in low density regions, as unveiled by the pioneering work of \citet{Vulcani+21}, only possible with IFU data, then this will introduce a bias in our results issued with Sample 2. The difficulty of selecting clean samples of RPS galaxies was also discussed by \citet{Laudari+22}.

If we look at the distribution of RPS candidates as a function of distance to the cluster centre (in units of $r_{200}$), the comparison of Fig.~\ref{fig:historsurr200} with Fig.~4 obtained for MACS0717 by \citet{Durret+21} shows that there are comparable, low numbers of RPS candidates near the cluster centres in MS0451 and MACS0717: RPS candidates 
tend to avoid the central, denser regions ($r<0.4\times r_{200}$) in both clusters. This was also the case in 97 RPS candidates in a sample of 39 clusters analyzed by \citet{Durret+21} -- see their Figure~6.
Hydrodynamical simulations have indeed found that RPS galaxies tend to avoid cluster centres \citep{Yun+19}.
This is different from what was observed at low redshift, where RPS galaxies were, on the contrary, in majority detected close to the central regions, for example by \cite{Jaffe+18} or \citet{Gullieuszik+20} - although these authors focus on galaxies with ionized gas tails, a very specific and extreme population of objects undergoing RPS. We acknowledge the fact that our redder selection band may miss such spectacular galaxies.

From the projected phase space diagrams of MS0451 and MACS0717 described in Section~\ref{sec:phase}, we can see that the distributions of galaxies between the virialized, infall, and backsplash regions differ between RPS and non-RPS galaxies, as well as from one cluster to the other. As seen in Table~\ref{tab:Mah}, in MS0451 66\% of the RPS candidates are virialized (with the definition $r<r_{100}$) while 79\% of the non-RPS galaxies are virialized. For both types of galaxies, there are more objects in the backsplash than in the infall region, and the difference is larger for RPS candidates. On the other hand, in MACS0717 only about half of all galaxies are virialized (49\% and 62\% for RPS and non-RPS galaxies), and the rest are distributed with 31\% in the infall region and 20\% in the backsplash region for RPS galaxies, and 
24\% in the infall region and 13\% in the backsplash region for non-RPS galaxies. These differences could be linked to the different dynamical histories of the two clusters.
Though \citet{Roberts+22} define virialized objects in a different way,
they mention that they see no evidence for stripping on first infall in their sample. This result seems to disagree with \cite{Jaffe+18}, who concluded from their study that 
many of the jellyfish galaxies seen in clusters likely formed via fast ($\sim 1-2$ Gyr) outside-in ram-pressure
stripping during their first infall into the cluster following highly radial orbits. 
This does not seem to be the case for MS0451, where the percentage of RPS candidates in the infall region is small, but is more likely to be true for MACS0717, where the percentage of RPS candidates in the infall region is higher.

 As discussed above, the plot of the SFR as a function of stellar mass (Fig.~\ref{fig:mass_SFR}) shows that about half of our RPS candidates are star-forming objects, lying within or above the  \cite{Vulcani+16} sequence, in contrast with non-RPS counterparts (of which the majority are located in the quiescent region of the plane).
 This result is comparable to that of \citet{Durret+21} for MACS0717, where an even stronger trend was found. Among the two clusters, MS0451 star-forming RPS galaxies show increased levels of SF activity, lying mostly above the MS. We can only conjecture that this is due to the different dynamical states of the clusters.
The enhanced median values of SFRs and sSFRs of our RPS candidates agree with those already found by previous authors such as \citet{Durret+21}, \citet{Roberts+22} and references therein.

Both clusters, MACS0717 and MS0451, located at z$\sim0.5$, host a non negligible population of RPS candidates showing important levels of star formation activity. It would be very interesting to see if dynamically "quieter/older" clusters, located at a similar redshift such as the four clusters observed by \cite{Cooper+22}, show similar results.
It has indeed been suggested \citep[e.g.][]{Owers+12,McPartland+16,Vulcani+17,EbelingKalita19,Moretti+22} that extreme ICM conditions in merging clusters tend to boost RPS events (that, in turn, seem to correlate with increased SFR as we have shown).

However, and in spite of this global trend, our MS0451 data and analysis points to a different behaviour for the low mass RPS candidates in this cluster: the ones we sample - in a notably incomplete way - below stellar masses of about $10^{9.4}$ M$_\odot$ have suppressed SF instead. This is a tentative result that apparently does not align with what has been found at lower redshifts and in different environments (non merging clusters and field), and strongly needs confirmation with better data and a finer analysis.

Finally: it is interesting to note that RPS candidates selected from the photometric redshift sample have properties that are comparable to those of our 56 spectroscopically confirmed RPS candidates, though they reach lower masses. This means that large samples of ram pressure stripped galaxies in clusters can potentially be obtained only based on photometric redshifts. With the advent of very large imaging surveys (DES, Euclid, LSST, etc.), this is a very promising method to acquire large samples of such objects at relatively low cost, and therefore to be able to study their physical properties in a statistical way. Of course, it will not be possible to look individually at
tens or hundreds of thousands of galaxies. However, machine learning should allow the automatic detection of asymmetries and particular features in galaxies in a near future, and by combining several constraints on the features expected from RPS galaxies, it should then become possible to extract large samples of RPS candidates, and thus to understand better the underlying physical processes that drive them. Simpler methods such as applying the scattering transform, as proposed by \cite{Cheng+21} are also very promising (B.~M\'enard, private communication).

\begin{acknowledgements}
  We are grateful to the referee for many constructive comments that helped us to improve the paper.
  We thank Andrea Biviano for enlightening discussions on phase space diagrams and for making some of his statistical programmes available to us.
  F.Durret acknowledges continuous support from CNES since 2002. 
  M. Jauzac is supported by the United Kingdom Research and Innovation (UKRI) Future Leaders Fellowship `Using Cosmic Beasts to uncover the Nature of Dark Matter' (grant number MR/S017216/1).
  H.Ebeling is grateful for financial support for programme GO-09722, provided by NASA through a grant from the Space Telescope Science Institute, which is operated by the Association of Universities for Research in Astronomy, Inc., under NASA contract NAS 5-26555.
  C. Lobo acknowledges support by Funda\c c\~ao para a Ci\^encia e a Tecnologia (FCT) through the research grants UIDB/04434/2020 and UIDP/04434/2020. C. Lobo further acknowledges support from the project "Identifying the Earliest Supermassive Black Holes with ALMA (IdEaS with ALMA)" (PTDC/FIS-AST/29245/2017). 
S.-I. Tam acknowledges support from the Ministry of Science and Technology of Taiwan (MOST grants 109-2112-M-001-018-MY3) and from the Academia Sinica Investigator Award (grant AS-IA-107-M01).

This work is partly based on tools and data products produced by GAZPAR operated by CeSAM-LAM and IAP and we further acknowledge the dedicated support of O. Ilbert.
This research has made use of the SVO Filter Profile Service (http://svo2.cab.inta-csic.es/theory/fps/) supported from the Spanish MINECO through grant AYA2017-84089.
This research has also made use of the NASA/IPAC Extragalactic Database (NED), which is funded by the National Aeronautics and Space Administration and operated by the California Institute of Technology.
We further acknowledge M. Taylor for developing TOPCAT, making our life so much easier.

\end{acknowledgements}

\bibliographystyle{aa}
\bibliography{sample.bib}

\begin{appendix}

\section{Results of the Zooniverse test}

\begin{table*}[h!]
\centering
\caption{Zooniverse classification of the 56 Sample 1 RPS candidates in MS0451. The columns are: running number, J classifications J$_L$ and J$_F$ according to two of us (LD and FD), mean J classification, number of times the galaxy was classified in Zooniverse, minimum J value in Zooniverse classifications, maximum J value in Zooniverse classifications, mean J value resulting from the Zooniverse classifications, dispersion on this J value, reason(s) to classify as an RPS candidate.}
\begin{tabular}{rrrrrrrrrl}
\hline \hline
Number & J$_L$ & J$_F$ & J$_{mean}$ & ${\rm n_{class}}$ & J$_{min}$ & J$_{max}$ & J$_{zoo}$ & $\sigma _{J_{zoo}}$ & comment  \\ 
\hline
 1 &  5 & 4 & 4.5 & 14 & 1 & 5 & 3.1 & 1.4 & edge, tails\\
 2 &  3 & 3 & 3.0 & 13 & 0 & 3 & 1.4 & 0.9 & sharp edge, tails \\
 3 &  4 & 3 & 3.5 & 12 & 1 & 3 & 2.1 & 0.9 & sharp edge \\
 4 &  2 & 2 & 2.0 & 11 & 1 & 4 & 2.7 & 1.1 & edges \\
 5 &  1 & 2 & 1.5 & 13 & 0 & 5 & 1.8 & 1.6 & moderately sharp edge\\
 6 &  3 & 3 & 3.0 & 17 & 0 & 3 & 1.5 & 0.8 & edge \\
 7 &  1 & 1 & 1.0 & 15 & 0 & 4 & 1.7 & 1.3 & small tail\\
 8 &  5 & 5 & 5.0 & 17 & 1 & 5 & 4.1 & 1.3 & edge+long tail\\
 9 &  3 & 3 & 3.0 & 12 & 0 & 5 & 1.8 & 1.5 & sharp edge \\
10 &  5 & 5 & 5.0 & 12 & 1 & 5 & 3.3 & 1.4 & edge+tail\\
11 &  2 & 3 & 2.5 & 13 & 1 & 5 & 3.0 & 1.5 & blobs and tail\\
12 &  2 & 2 & 2.0 & 12 & 0 & 3 & 1.2 & 1.1 & tails \\
13 &  2 & 2 & 2.0 & 10 & 0 & 3 & 1.1 & 0.7 & bright knots \\
14 &  1 & 1 & 1.0 & 16 & 0 & 3 & 2.2 & 0.9 & tail \\
15 &  2 & 3 & 2.5 & 15 & 0 & 5 & 2.7 & 1.1 & tail \\
16 &  2 & 2 & 2.0 & 15 & 0 & 5 & 2.3 & 1.4 & edges, very disturbed\\
17 &  2 & 1 & 1.5 & 15 & 0 & 4 & 1.3 & 1.2 & edges, tails \\
18 &  1 & 1 & 1.0 & 16 & 0 & 5 & 2.5 & 1.4 & edges, tail\\
19 &  3 & 4 & 3.5 & 14 & 0 & 5 & 2.8 & 1.6 & edge, tail \\
20 &  2 & 2 & 2.0 & 13 & 0 & 3 & 1.8 & 1.1 & sharp edge \\
21 &  3 & 3 & 3.0 & 14 & 0 & 3 & 1.9 & 1.2 & edge, tail \\
22 &  1 & 2 & 1.5 & 13 & 0 & 4 & 2.4 & 1.3 & tail \\
23 &  1 & 1 & 1.0 & 14 & 0 & 2 & 0.7 & 0.6 & moderately sharp edge \\
24 &  1 & 1 & 1.0 & 11 & 0 & 3 & 1.4 & 0.8 & moderately sharp edges \\
25 &  3 & 3 & 3.0 & 14 & 0 & 2 & 1.1 & 0.8 & edges \\
26 &  1 & 2 & 1.5 & 14 & 0 & 3 & 1.4 & 1.1 & tail \\
27 &  5 & 5 & 5.0 & 12 & 2 & 5 & 3.8 & 0.8 & edges+tail \\
28 &  2 & 3 & 2.5 & 14 & 0 & 3 & 1.6 & 0.7 & tails \\
29 &  1 & 2 & 1.5 & 14 & 0 & 4 & 1.4 & 1.1 & moderately sharp edge \\
30 &  2 & 3 & 2.5 & 16 & 0 & 5 & 3.4 & 1.5 & edges \\ 
31 &  1 & 1 & 1.0 & 12 & 0 & 3 & 1.2 & 0.7 & asymmetric+edge\\
32 &  3 & 3 & 3.0 & 13 & 0 & 4 & 2.0 & 1.2 & edge, blobs \\
33 &  1 & 1 & 1.0 & 13 & 0 & 3 & 0.8 & 0.8 & diffuse, bent \\
34 &  1 & 1 & 1.0 & 14 & 0 & 2 & 0.6 & 0.7 & weak tail \\
35 &  5 & 5 & 5.0 & 12 & 2 & 5 & 4.2 & 1.0 & sharp edges \\
36 &  1 & 1 & 1.0 & 12 & 0 & 5 & 1.9 & 1.4 & tails\\
37 &  2 & 2 & 2.0 & 13 & 0 & 5 & 2.0 & 1.6 & edges +blobs\\
38 &  5 & 5 & 5.0 & 13 & 1 & 5 & 3.1 & 1.4 & sharp edges+blobs\\
39 &  2 & 2 & 2.0 & 18 & 0 & 5 & 2.6 & 1.2 & edge \\
40 &  3 & 3 & 3.0 & 11 & 1 & 4 & 3.0 & 1.1 & tail \\
41 &  2 & 2 & 2.0 & 16 & 0 & 3 & 1.6 & 1.0 & weak tail \\
42 &  2 & 2 & 2.0 & 13 & 1 & 3 & 2.0 & 0.8 & tail \\
43 &  1 & 1 & 1.0 & 14 & 1 & 4 & 2.4 & 1.0 & edge \\
44 &  1 & 1 & 1.0 & 14 & 0 & 5 & 2.3 & 2.0 & edges \\
45 &  1 & 1 & 1.0 & 16 & 0 & 4 & 2.4 & 1.2 & edge, tail \\
46 &  1 & 2 & 1.5 & 15 & 0 & 4 & 1.8 & 1.3 & edge \\
47 &  1 & 2 & 1.5 & 16 & 1 & 5 & 3.0 & 1.4 & edge, possible tail \\
48 &  3 & 3 & 3.0 & 14 & 1 & 5 & 3.0 & 1.2 & edge \\
49 &  1 & 1 & 1.0 & 12 & 0 & 4 & 1.9 & 1.4 & possible tail \\
50 &  1 & 1 & 1.0 & 15 & 1 & 4 & 2.1 & 1.0 & edge, possible tail \\
51 &  1 & 1 & 1.0 & 14 & 1 & 3 & 2.0 & 0.7 & possible tail \\
52 &  1 & 1 & 1.0 & 13 & 0 & 2 & 0.9 & 0.9 & possible tail \\
53 &  2 & 2 & 2.0 & 13 & 0 & 5 & 2.2 & 1.5 & tails \\
54 &  3 & 3 & 3.0 & 15 & 2 & 4 & 3.0 & 0.8 & sharp edge, tail \\
55 &  1 & 2 & 1.5 & 12 & 1 & 4 & 2.2 & 1.1 & weak tail \\
56 &  2 & 2 & 2.0 & 13 & 1 & 4 & 2.5 & 1.1 & tail \\
\hline
\end{tabular}
\label{tab:zoo}
\end{table*}

\begin{figure*}[h!]
\begin{center}
\includegraphics[width=0.9\textwidth]{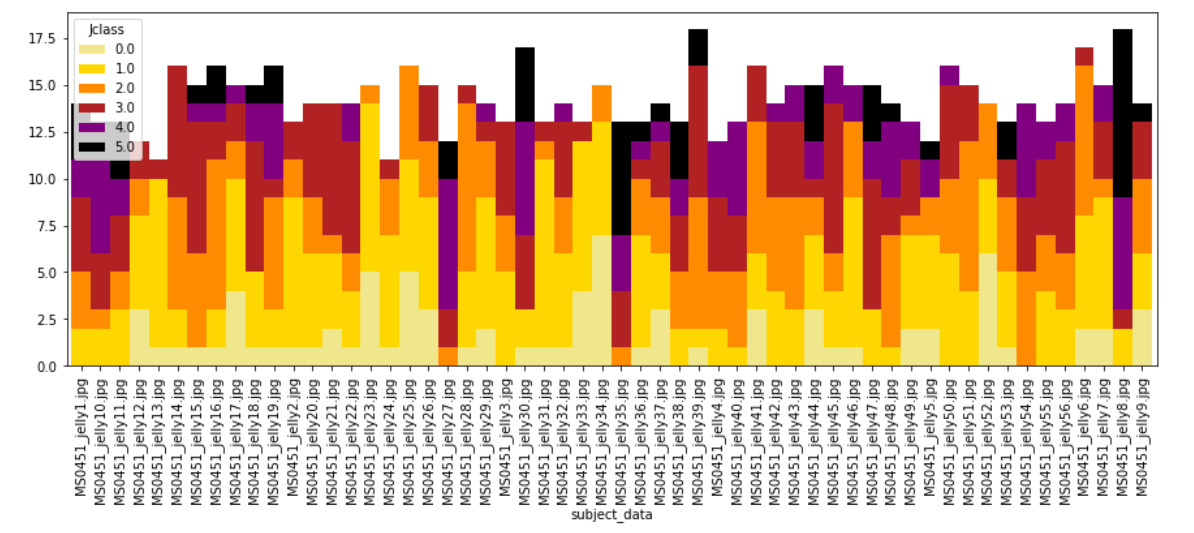} 
\end{center}
\caption{Grades given by the Zooniverse classifiers for each of the 56 galaxies of Sample~1.
}
\label{fig:Zoo_histoJ}
\end{figure*}

A summary of the results of the J classification with Zooniverse is shown in Fig.~\ref{fig:Zoo_histoJ} and Table~\ref{tab:zoo}. The corresponding results are discussed in Section~\ref{sec:zoo}.


\section{Results of KS tests}

\begin{table*}[h!]
\centering
\caption{Percentages at which a KS test does not reject the hypothesis that the
two samples are drawn from the same parent population. The second and third columns give the values for the whole sample, the fourth column gives the corresponding values after applying a mass cut at $10^9\leq$M$\leq 10^{10.5}$~M$_\odot$.} 
\begin{tabular}{llll}
\hline \hline
Quantity & samples & percentages & percentages \\
\hline
Masses  & S1-S3 & 31.6\% & 28.5\% \\
        & S2-S4 & 27.1\% & 16.1\% \\
Ages    & S1-S3 & 73.6\% & 84.9\% \\
        & S2-S4 & 47.7\% & 47.8\% \\
SFR	    &S1-S3  & 43.9\% & 43.7\% \\
        &S2-S4  & 31.2\% & 43.1\% \\
sSFR    &S1-S3  & 57.6\% & 46.3\% \\
        &S2-S4  & 46.5\% & 47.5\% \\
\hline
\end{tabular}
\label{tab:KS}
\end{table*}

The results of the KS tests applied in Section~4 are given in Table~\ref{tab:KS}.

\section{Images of RPS galaxies in MS0451}
 \label{appendix1}

The images of the 56 RPS galaxy candidates with spectroscopic redshifts in MS0451
are shown below. For each galaxy, we indicate the classifications estimated by two of us in parentheses (as in Table~\ref{tab:jelly}).

\begin{figure*}[h]
\begin{center}
\caption{From top to bottom and left to right, we show galaxies \#1 (J-class 5-4), \#2 (3-3), \#3 (4-3), \#4 (2-2), \#5 (1-2), \#6 (3-3), \#7 (1-1), \#8 (5-5), \#9 (3-3), \#10 (5-5), \#11 (2-3), and \#12 (2-2) in the F814W filter.}
  \includegraphics[width=5.5cm,height=5.5cm]{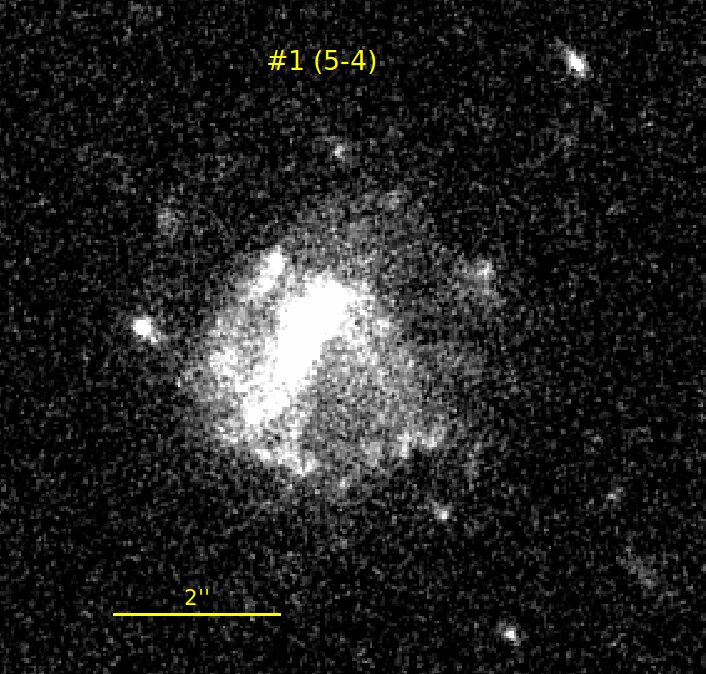} \kern0.1cm%
  \includegraphics[width=5.5cm,height=5.5cm]{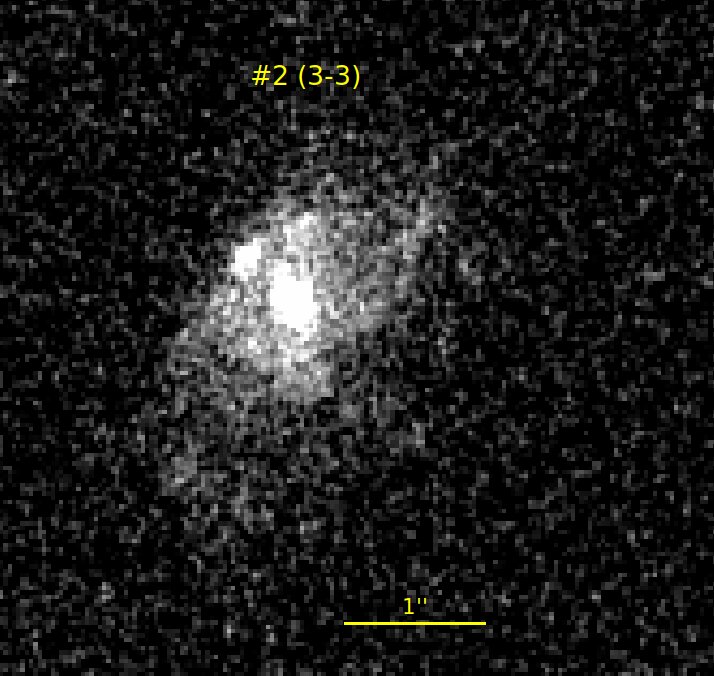} 
  \includegraphics[width=5.5cm,height=5.5cm]{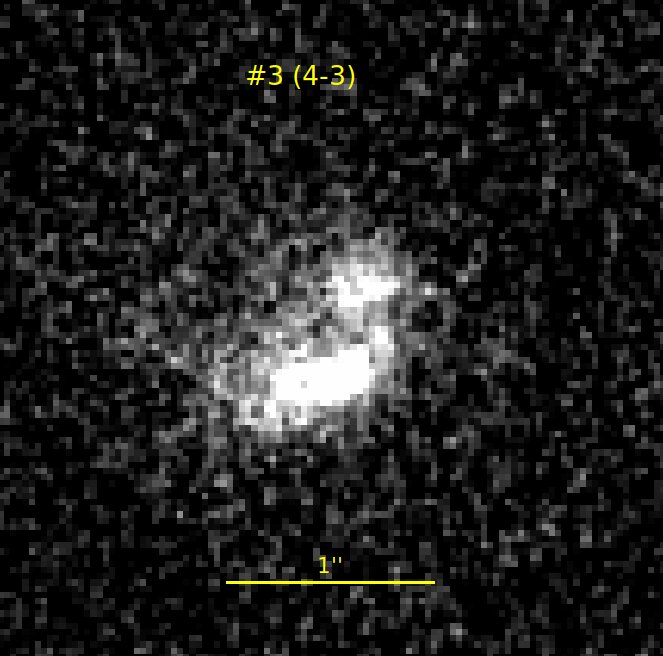}  \kern0.1cm
  \includegraphics[width=5.5cm,height=5.5cm]{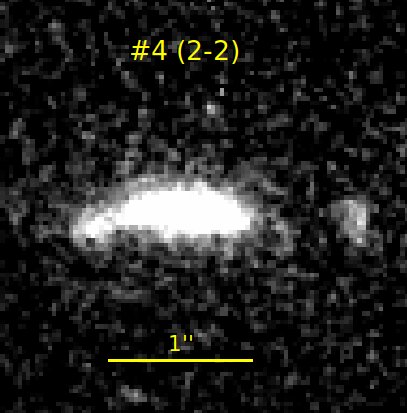}  
  \includegraphics[width=5.5cm,height=5.5cm]{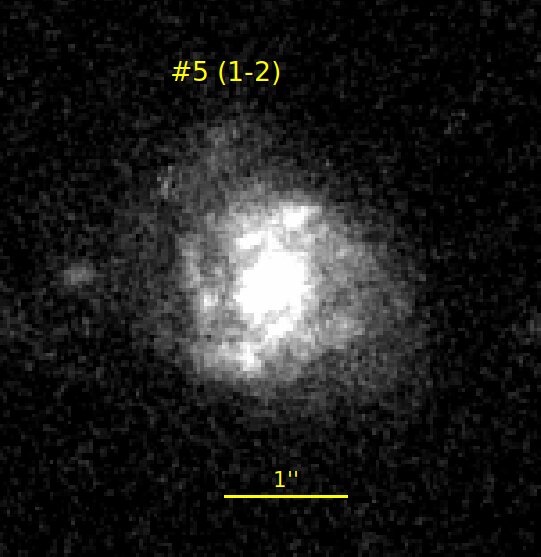} \kern0.1cm%
  \includegraphics[width=5.5cm,height=5.5cm]{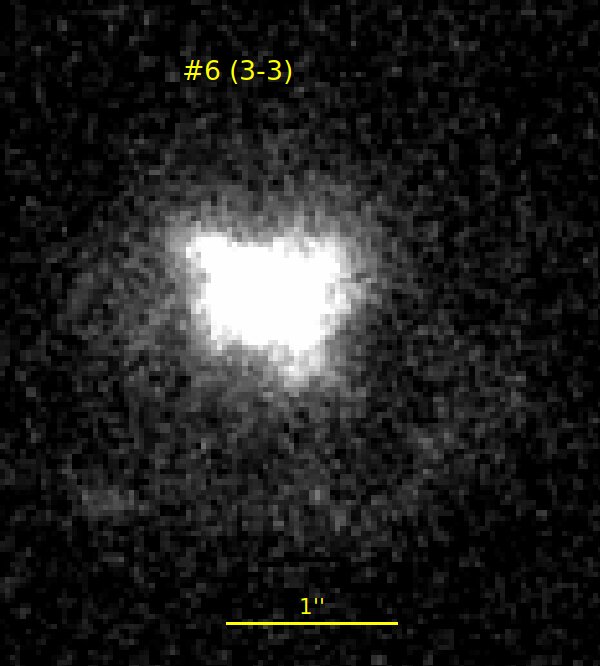}  
  \includegraphics[width=5.5cm,height=5.5cm]{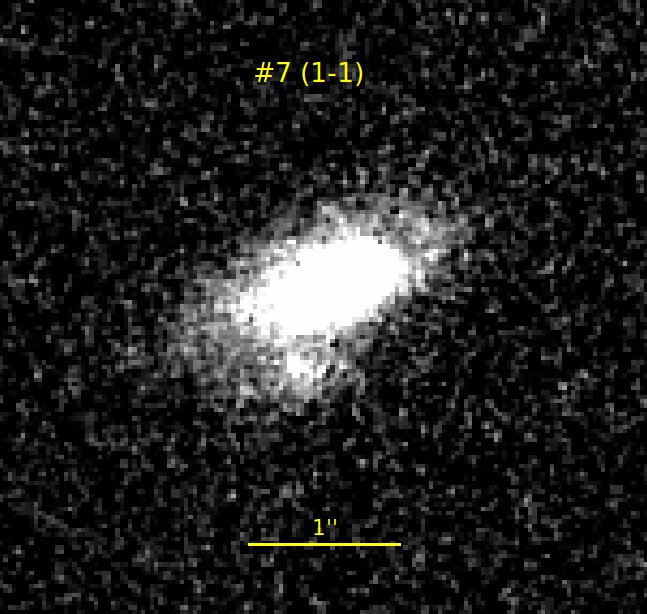} \kern0.1cm%
  \includegraphics[width=5.5cm,height=5.5cm]{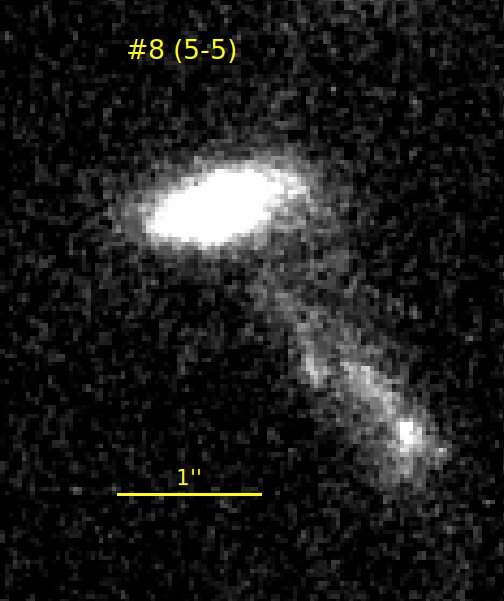} 
  \includegraphics[width=5.5cm,height=5.5cm]{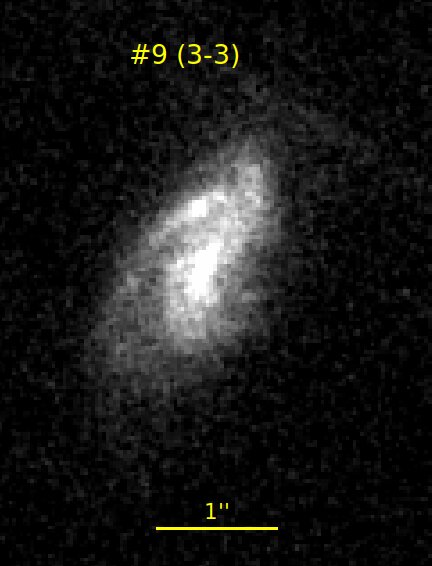}  \kern0.1cm%
  \includegraphics[width=5.5cm,height=5.5cm]{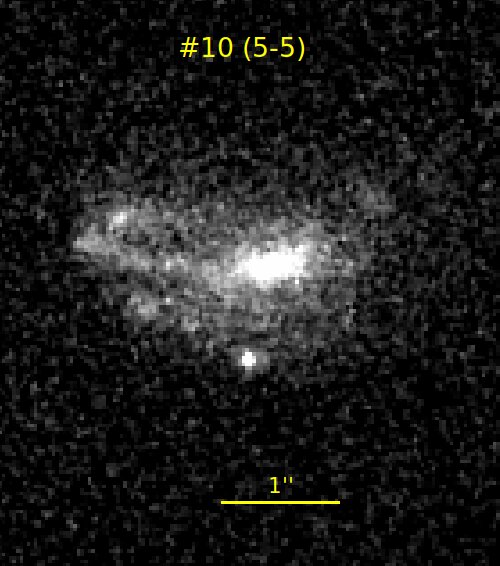} 
  \includegraphics[width=5.5cm,height=5.5cm]{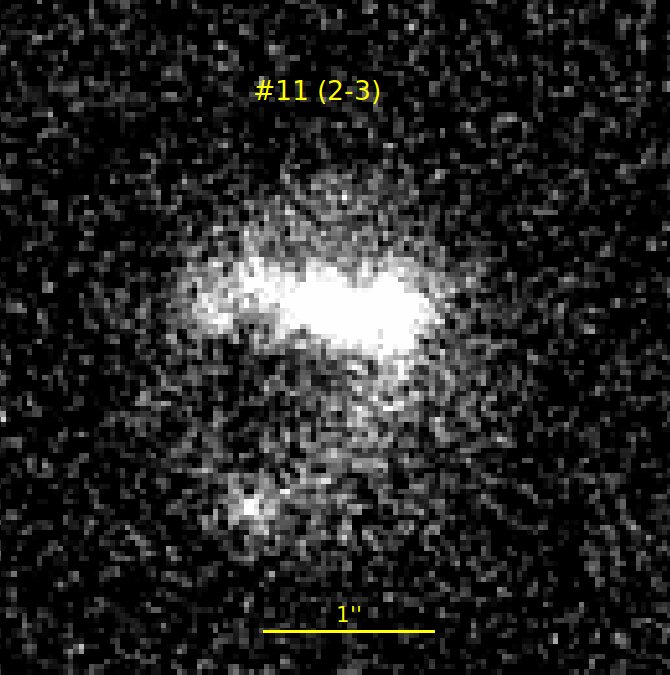}  \kern0.1cm
  \includegraphics[width=5.5cm,height=5.5cm]{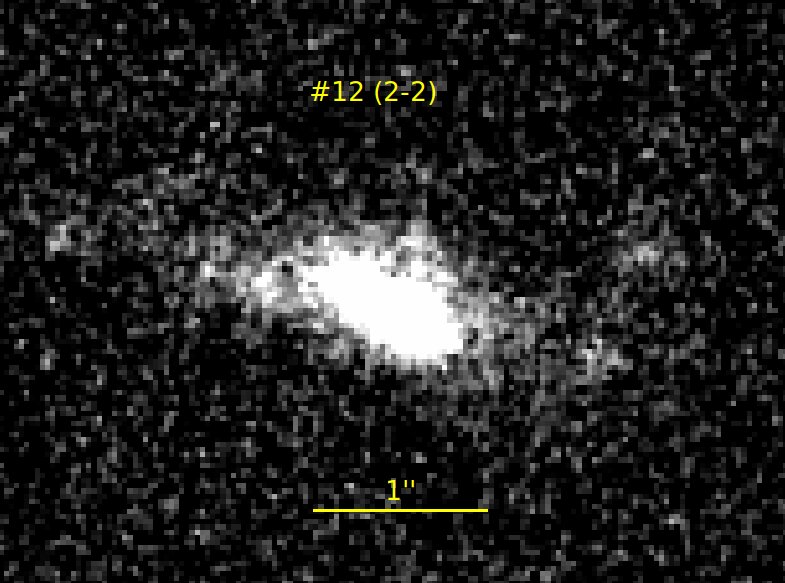} 
\end{center}
\end{figure*}

\begin{figure*}[h]
\begin{center}
\caption{From top to bottom and left to right, we show galaxies \#13 (2-2), \#14 (1-1), \#15 (2-3), \#16 (2-2), \#17 (2-1), \#18 (1-1), \#19 (3-4), \#20 (2-2), \#21 (3-3), \#22 (1-2), \#23 (1-1), and \#24 (1-1), in the F814W filter.}
  \includegraphics[width=5.5cm,height=5.5cm]{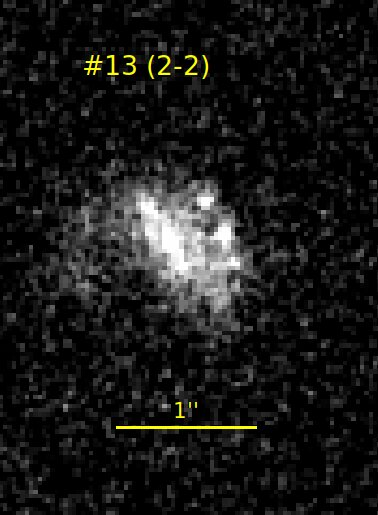} \kern0.1cm%
  \includegraphics[width=5.5cm,height=5.5cm]{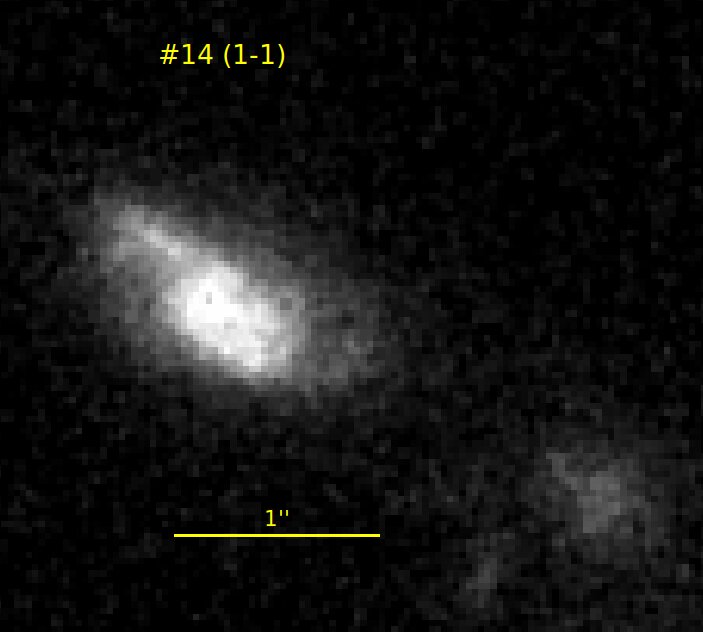} 
  \includegraphics[width=5.5cm,height=5.5cm]{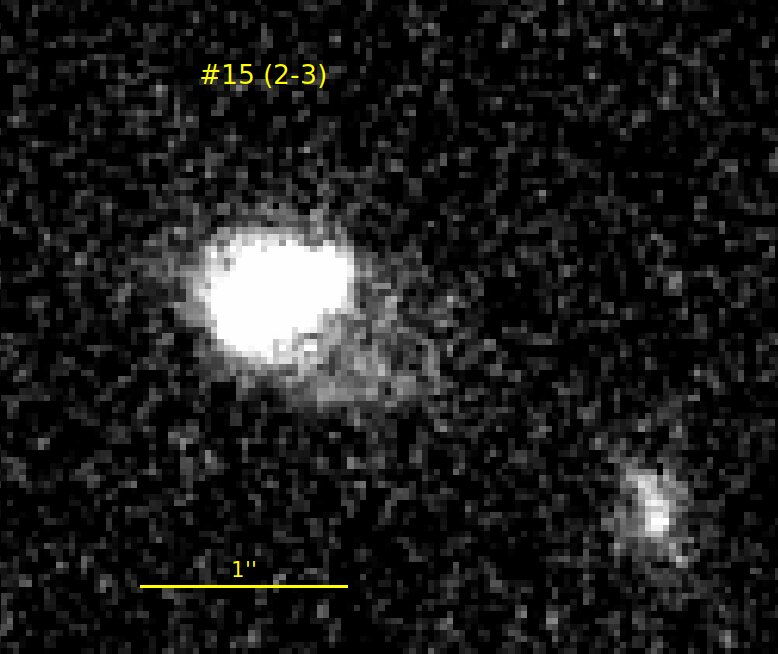}  \kern0.1cm
  \includegraphics[width=5.5cm,height=5.5cm]{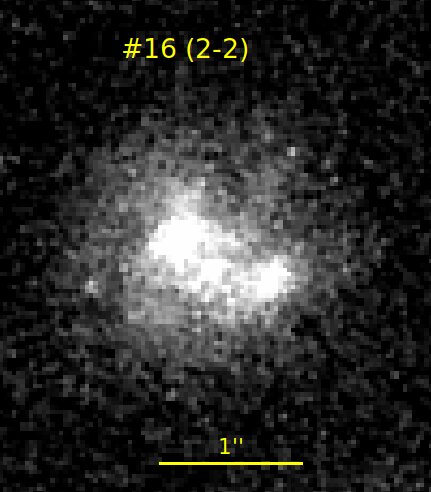}  
  \includegraphics[width=5.5cm,height=5.5cm]{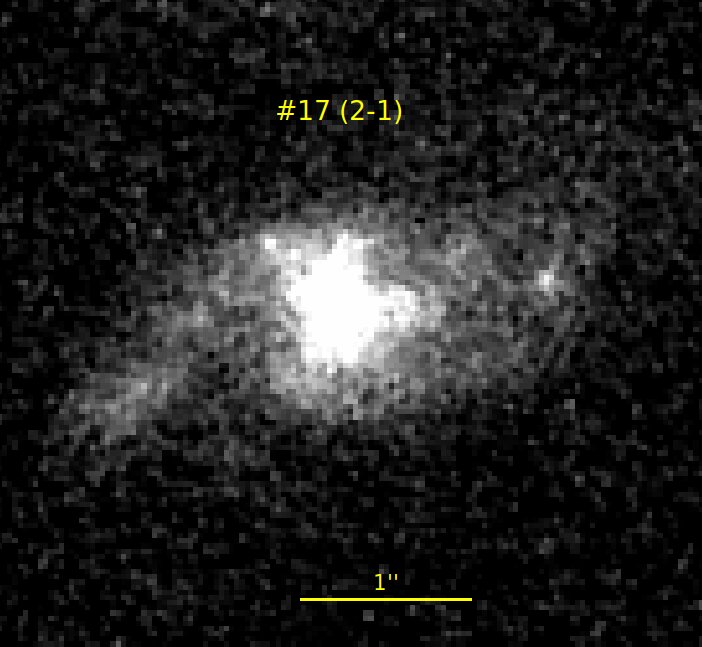} \kern0.1cm%
  \includegraphics[width=5.5cm,height=5.5cm]{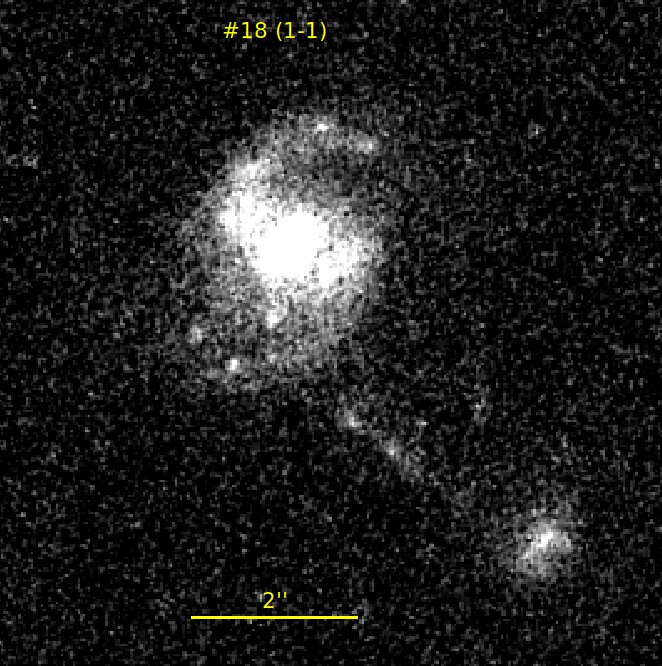}  
  \includegraphics[width=5.5cm,height=5.5cm]{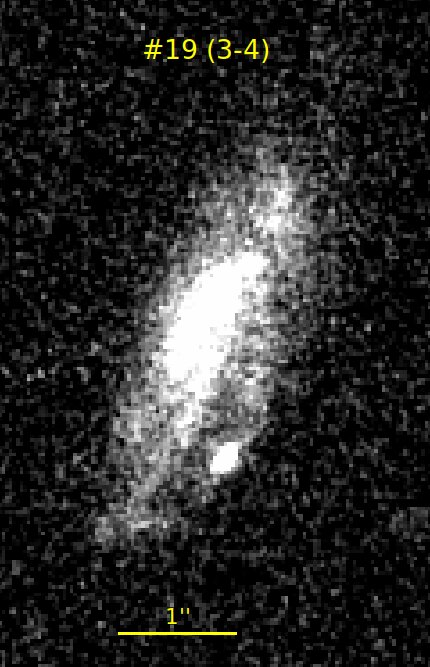} \kern0.1cm%
  \includegraphics[width=5.5cm,,height=5.5cm]{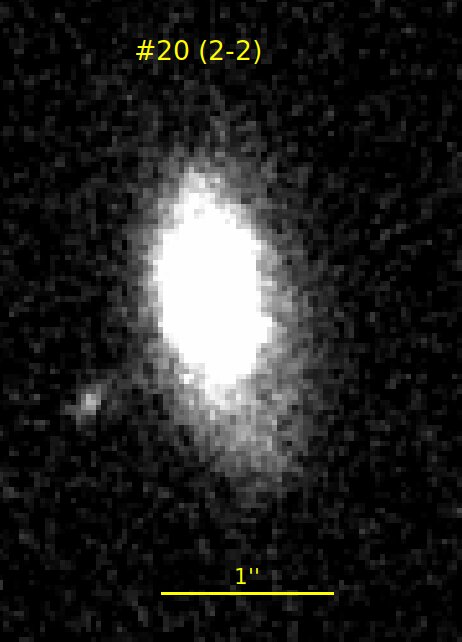} 
  \includegraphics[width=5.5cm,,height=5.5cm]{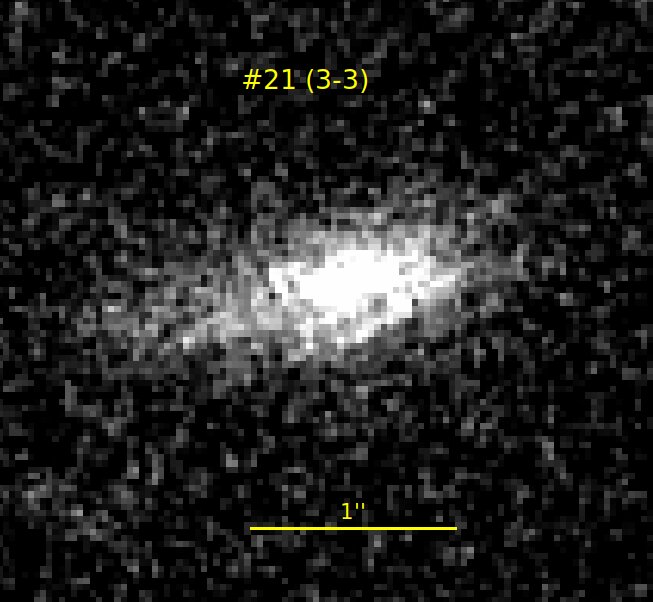}  \kern0.1cm%
  \includegraphics[width=5.5cm,height=5.5cm]{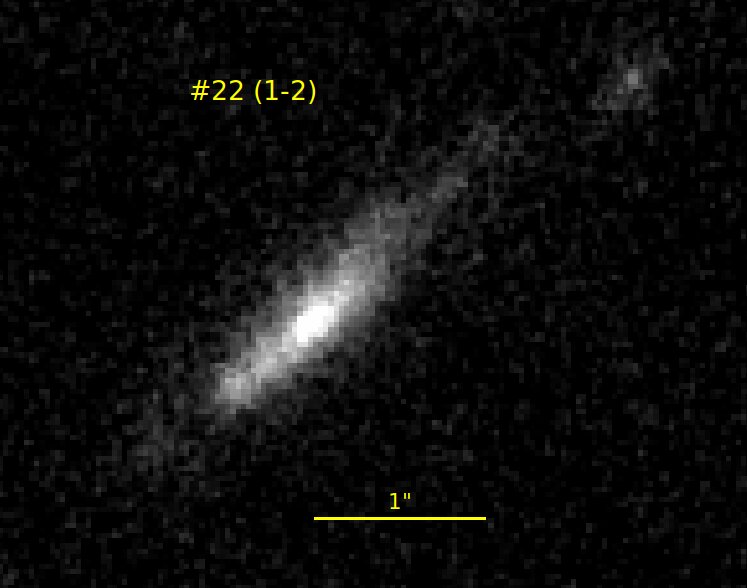} 
  \includegraphics[width=5.5cm,height=5.5cm]{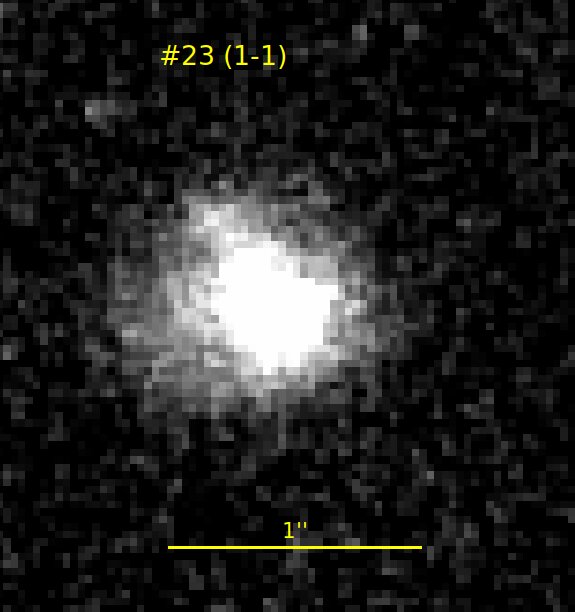}  \kern0.1cm
  \includegraphics[width=5.5cm,,height=5.5cm]{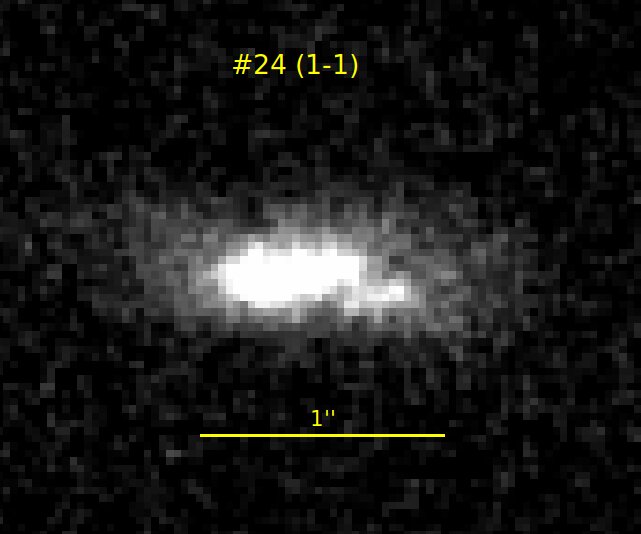} 
\end{center}
\end{figure*}

\begin{figure*}[h]
\begin{center}
\caption{From top to bottom and left to right we show galaxies \#25 (3-3), \#26 (1-2), \#27 (5-5), \#28 (2-3), \#29 (1-2), \#30 (2-3), \#31 (1-1), \#32 (3-3), \#33 (1-1), \#34 (1-1), \#35 (5-5), and \#36 (1-1), in the F814W filter.}
  \includegraphics[width=5.5cm,height=5.5cm]{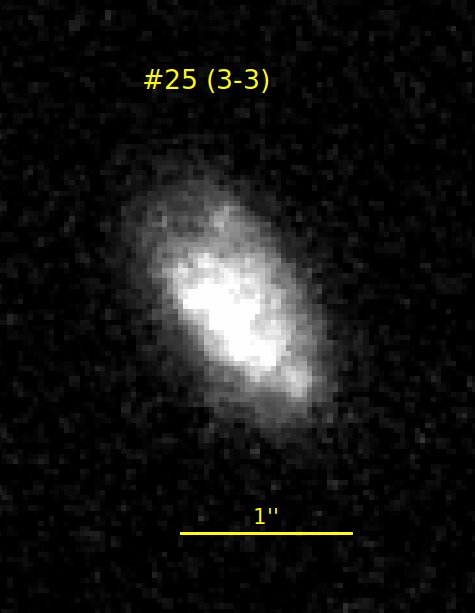} \kern0.1cm%
  \includegraphics[width=5.5cm,height=5.5cm]{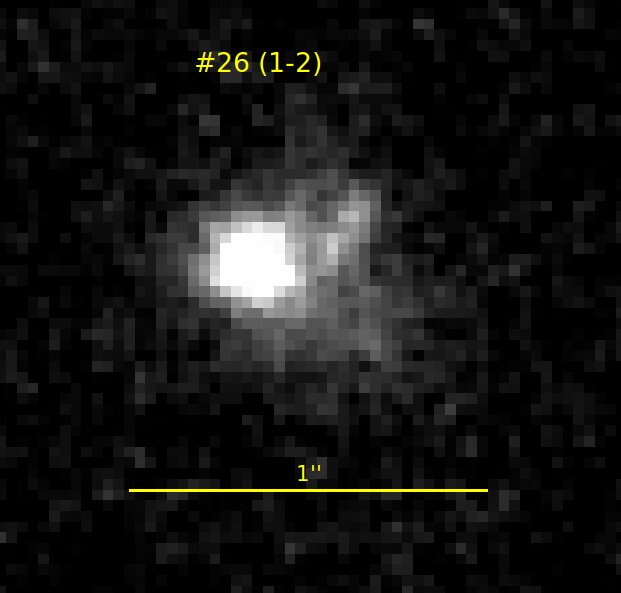}  
  \includegraphics[width=5.5cm,height=5.5cm]{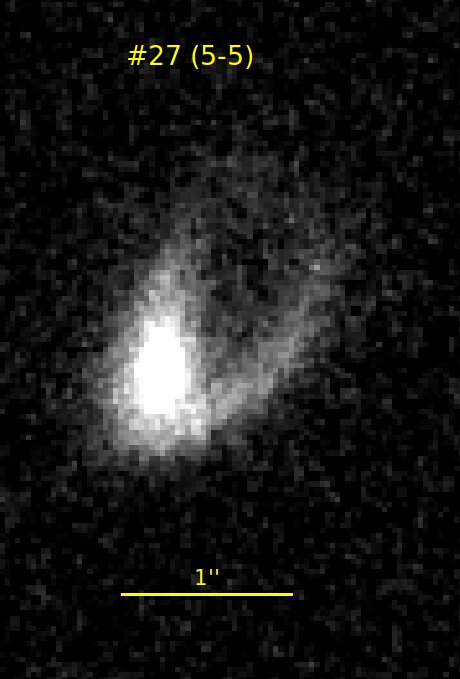} \kern0.1cm%
  \includegraphics[width=5.5cm,height=5.5cm]{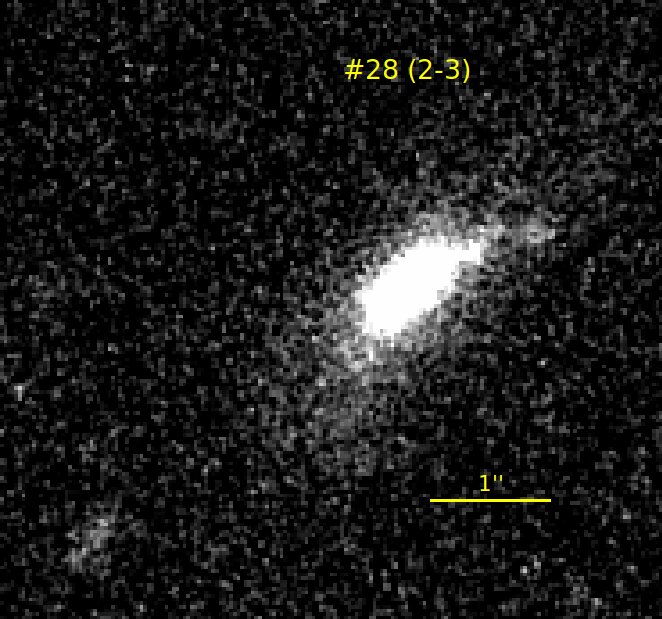}  
  \includegraphics[width=5.5cm,height=5.5cm]{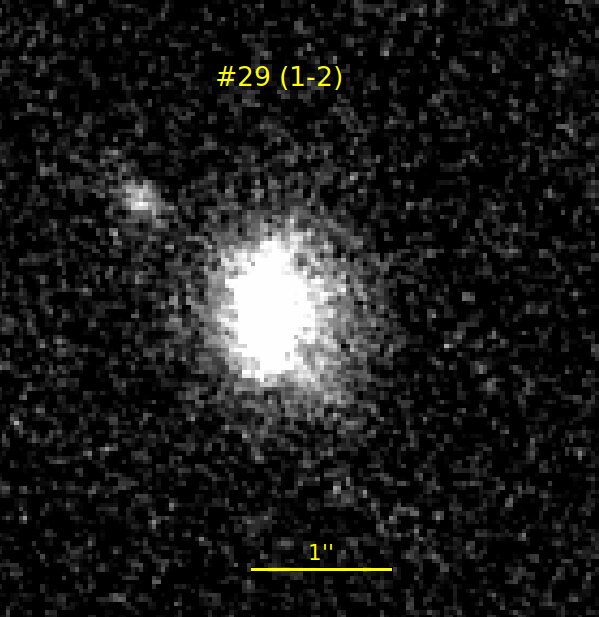} \kern0.1cm%
  \includegraphics[width=5.5cm,height=5.5cm]{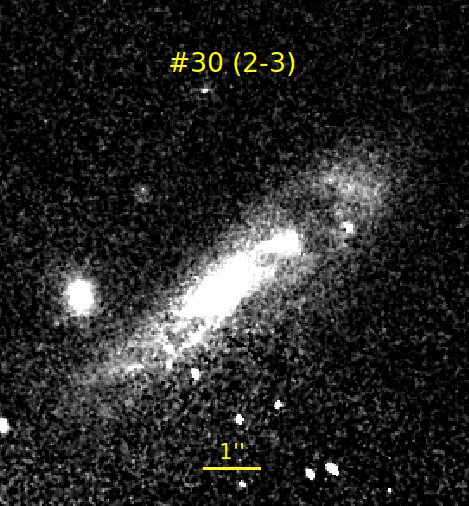} 
  \includegraphics[width=5.5cm,height=5.5cm]{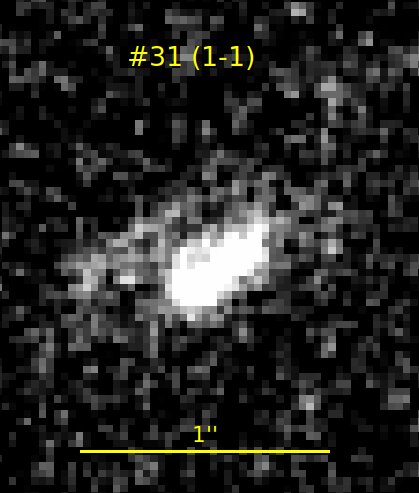} \kern0.1cm%
  \includegraphics[width=5.5cm,height=5.5cm]{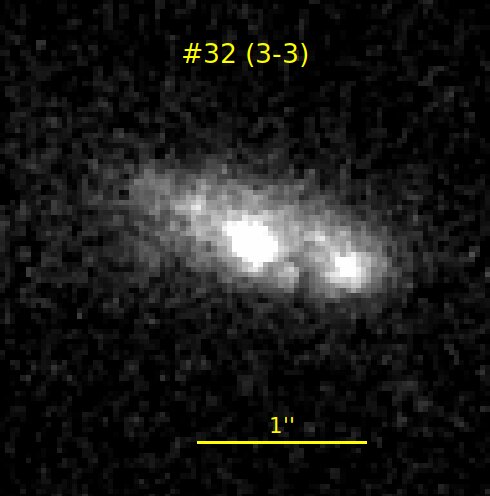} 
  \includegraphics[width=5.5cm,height=5.5cm]{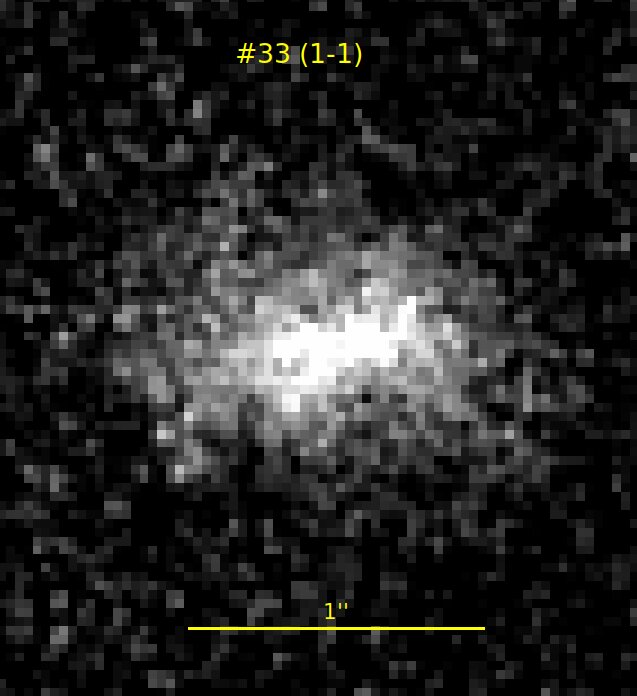} \kern0.1cm%
  \includegraphics[width=5.5cm,height=5.5cm]{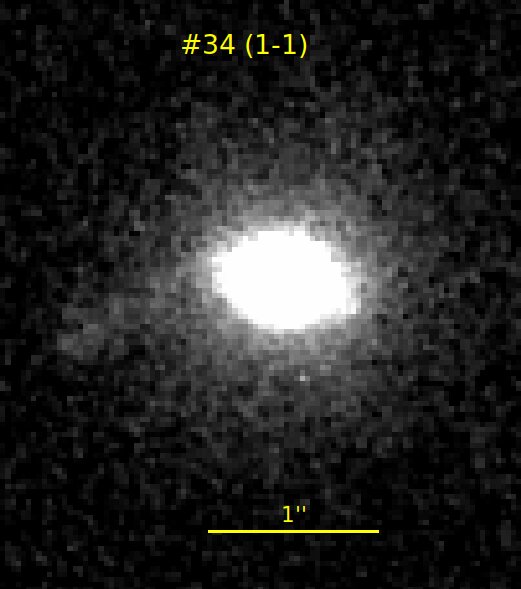} 
  \includegraphics[width=5.5cm,height=5.5cm]{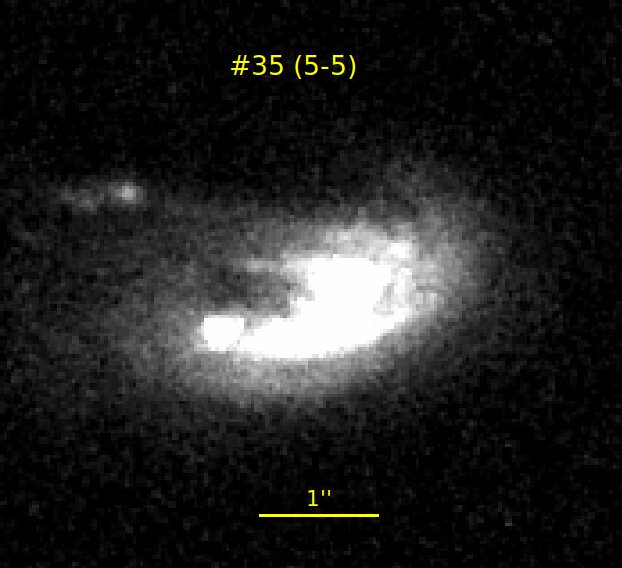} \kern0.1cm%
  \includegraphics[width=5.5cm,height=5.5cm]{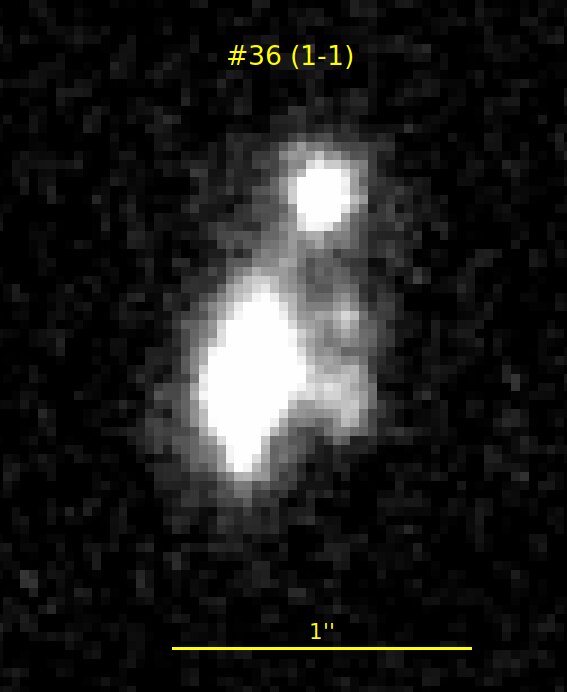}  
\end{center}
\end{figure*}

 \begin{figure*}[h]
\begin{center}
\caption{From top to bottom and left to right we show galaxies  \#37 (2-2), \#38 (5-5), \#39 (2-2), \#40 (3-3), \#41 (2-2), \#42 (2-2), \#43 (1-1), \#44 (1-1), \#45 (1-1), \#46 (1-2), \#47 (1-2), and \#48 (3-3), in the F814W filter.}
  \includegraphics[width=5.5cm,height=5.5cm]{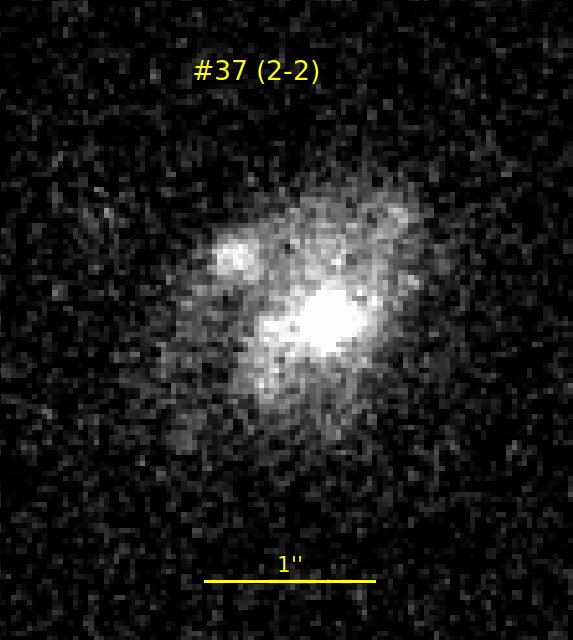} \kern0.1cm%
  \includegraphics[width=5.5cm,height=5.5cm]{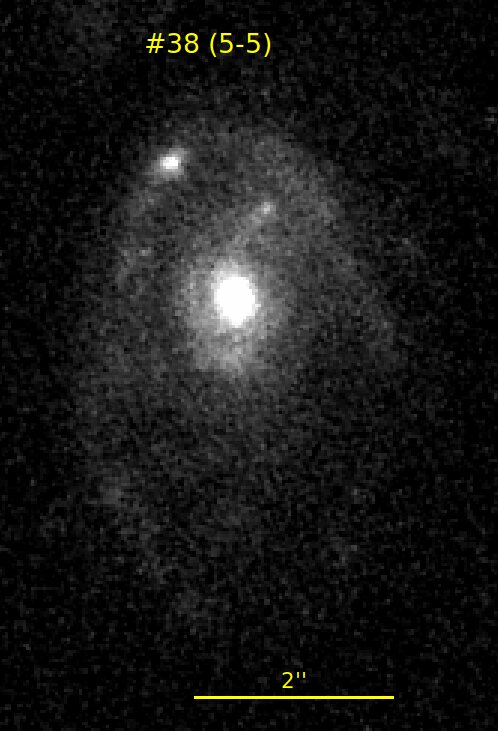}  
  \includegraphics[width=5.5cm,height=5.5cm]{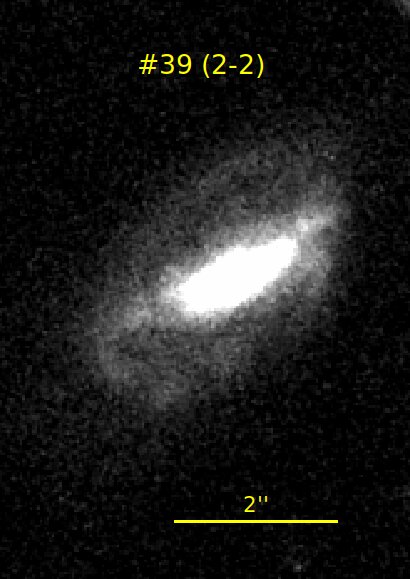} \kern0.1cm%
  \includegraphics[width=5.5cm,height=5.5cm]{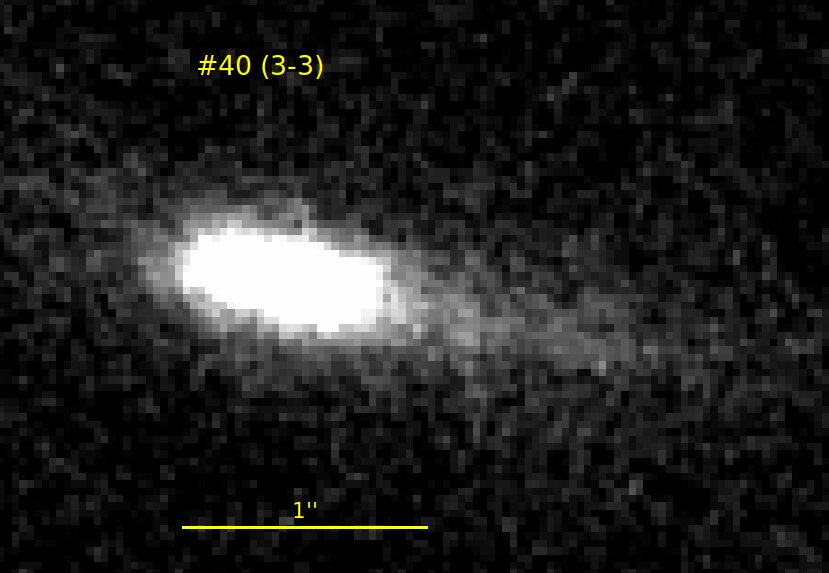} 
  \includegraphics[width=5.5cm,height=5.5cm]{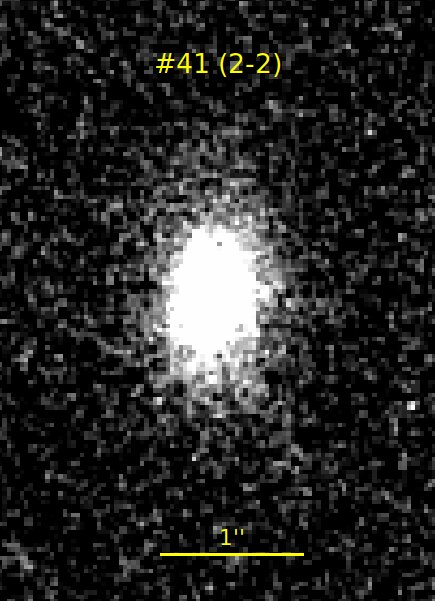} \kern0.1cm%
  \includegraphics[width=5.5cm,height=5.5cm]{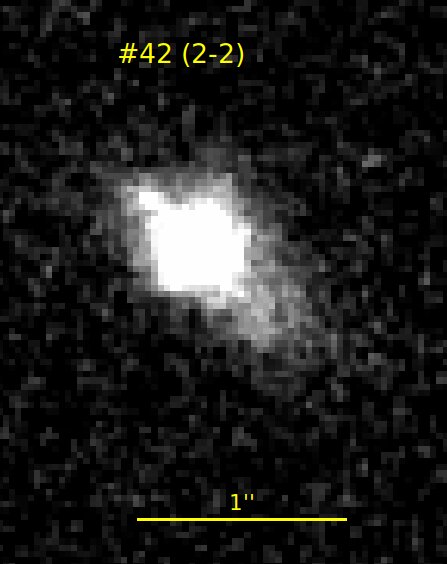} 
  \includegraphics[width=5.5cm,height=5.5cm]{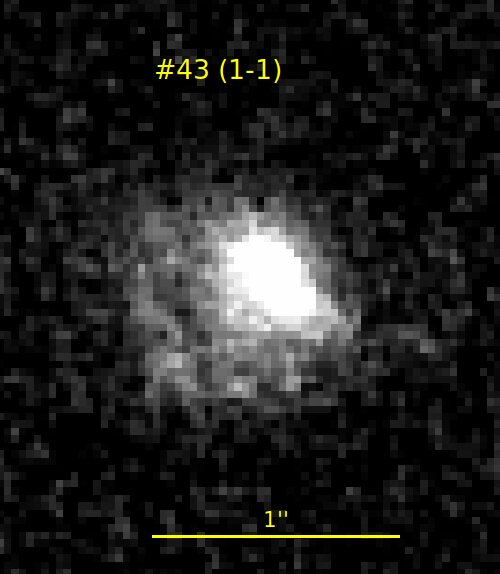} \kern0.1cm%
  \includegraphics[width=5.5cm,height=5.5cm]{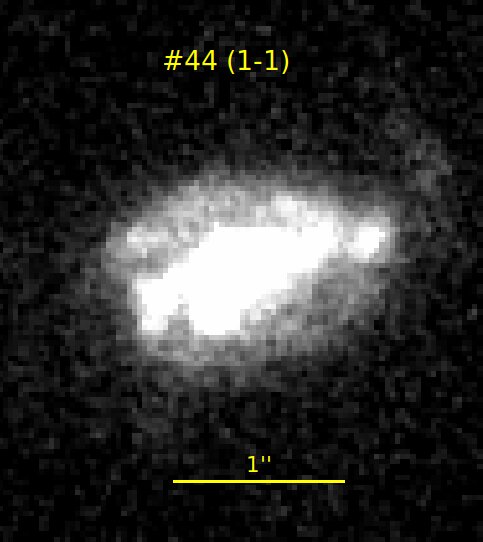} 
  \includegraphics[width=5.5cm,height=5.5cm]{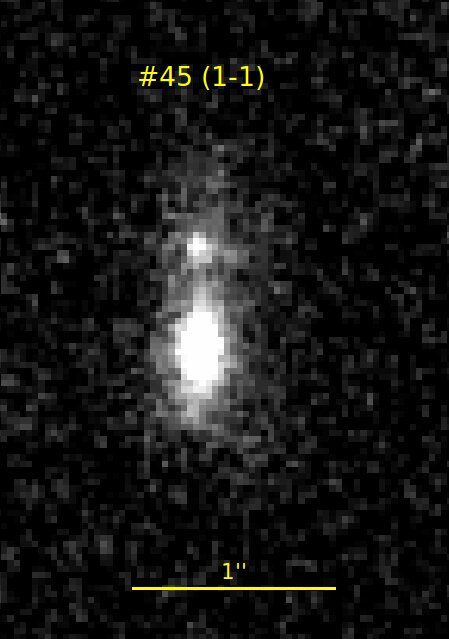} \kern0.1cm%
  \includegraphics[width=5.5cm,height=5.5cm]{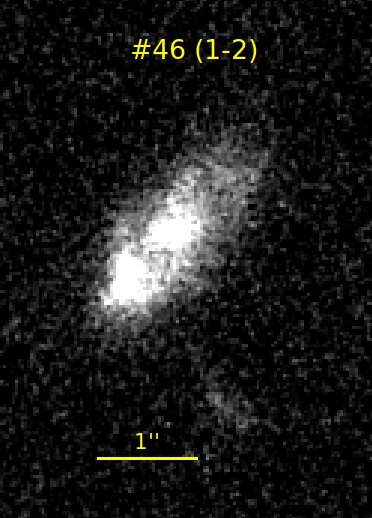}  
  \includegraphics[width=5.5cm,height=5.5cm]{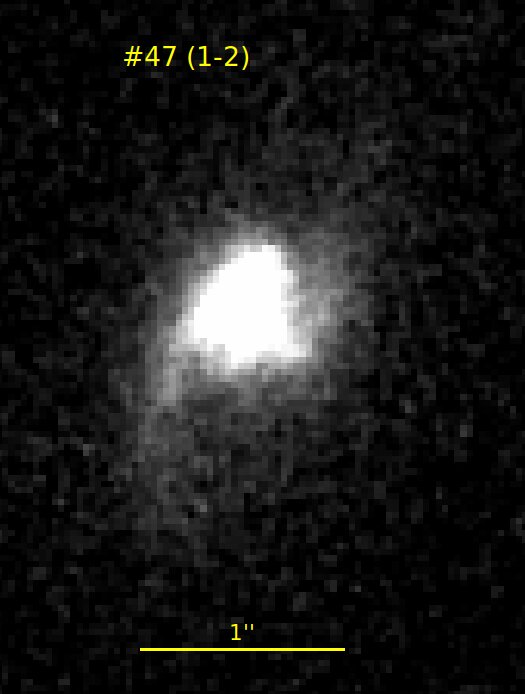} \kern0.1cm%
  \includegraphics[width=5.5cm,height=5.5cm]{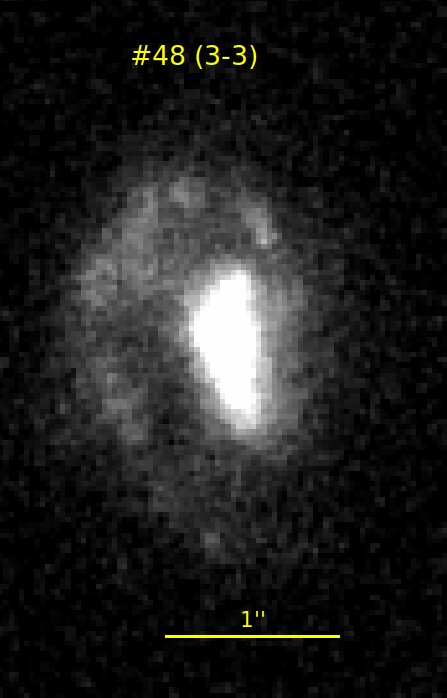}  
\end{center}
\end{figure*}

\begin{figure*}[h]
\begin{center}
\caption{From top to bottom and left to right we show galaxies  \#49 (1-1), \#50 (1-1), \#51 (1-1), \#52 (1-1), \#53 (2-2), \#54 (3-3), \#55 (1-2), and \#56 (2-2), in the F814W filter.}
  \includegraphics[width=5.5cm,height=5.5cm]{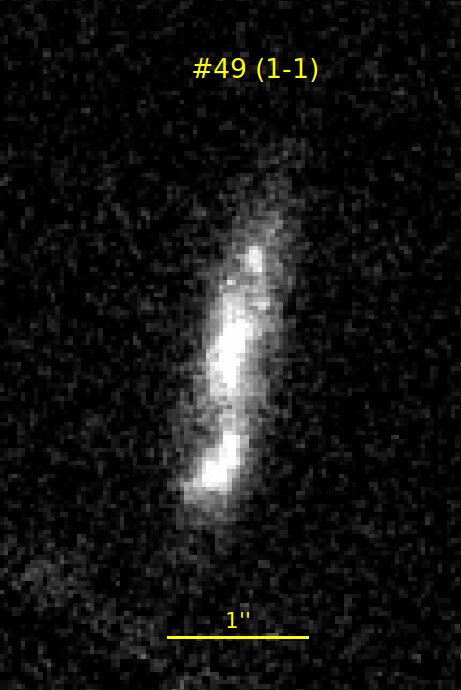} \kern0.1cm%
  \includegraphics[width=5.5cm,height=5.5cm]{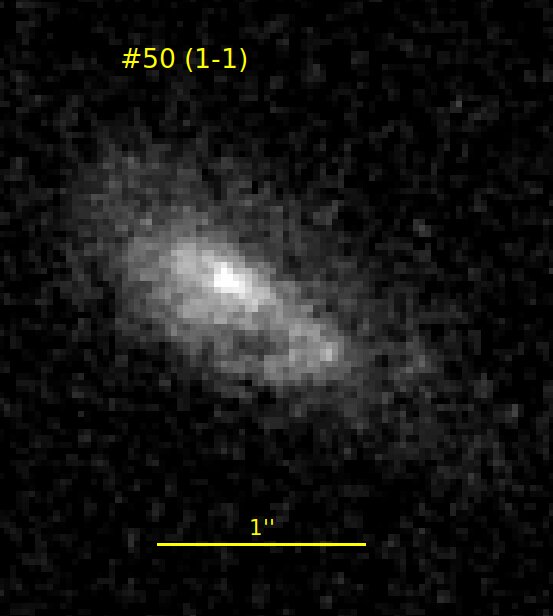} 
  \includegraphics[width=5.5cm,height=5.5cm]{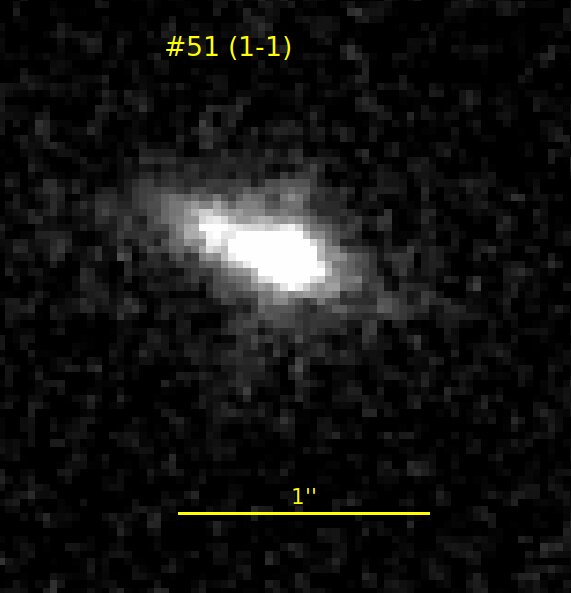} \kern0.1cm%
  \includegraphics[width=5.5cm,height=5.5cm]{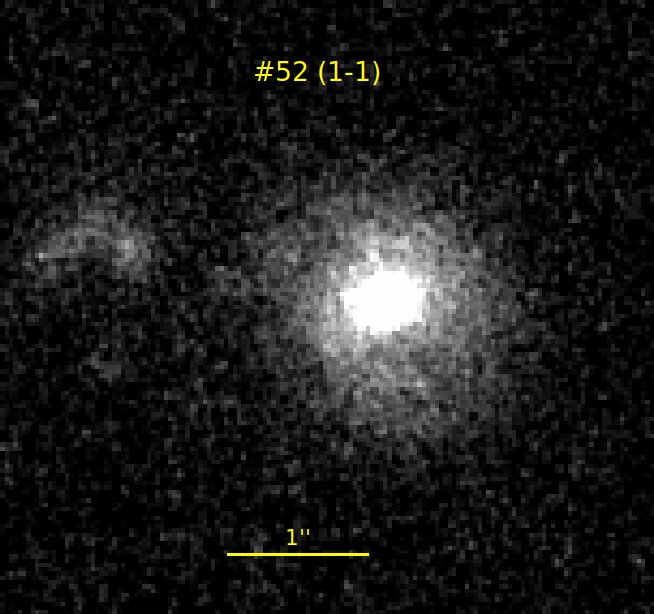} 
  \includegraphics[width=5.5cm,height=5.5cm]{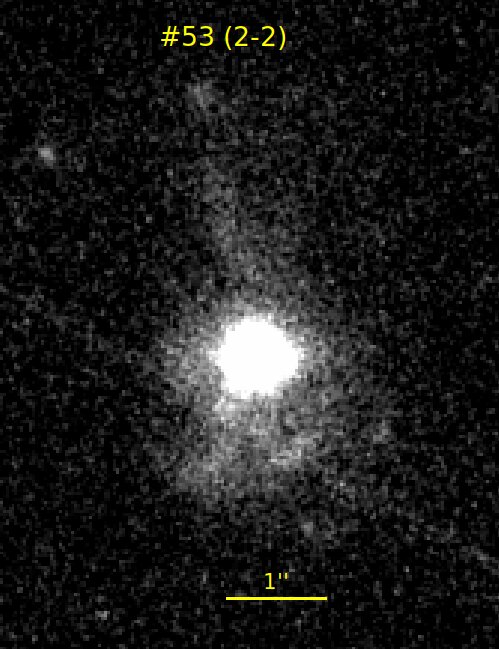} \kern0.1cm%
  \includegraphics[width=5.5cm,height=5.5cm]{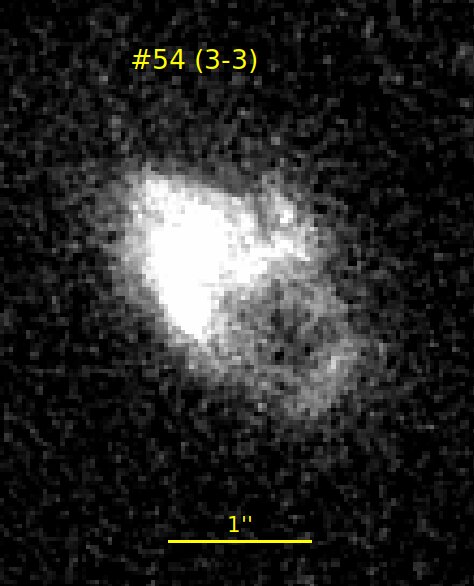} 
  \includegraphics[width=5.5cm,height=5.5cm]{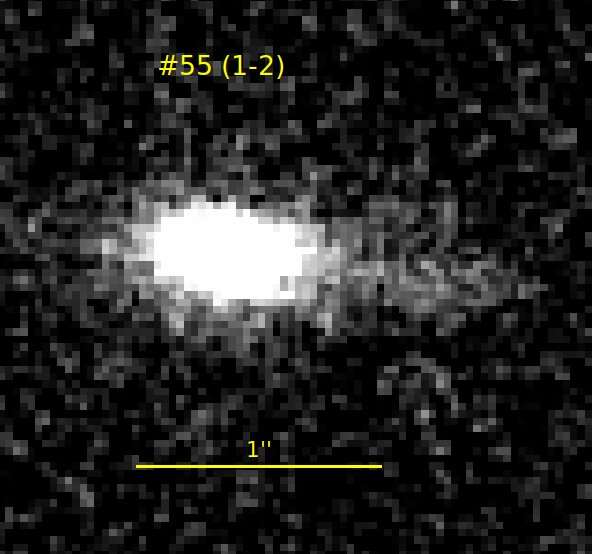} \kern0.1cm%
  \includegraphics[width=5.5cm,height=5.5cm]{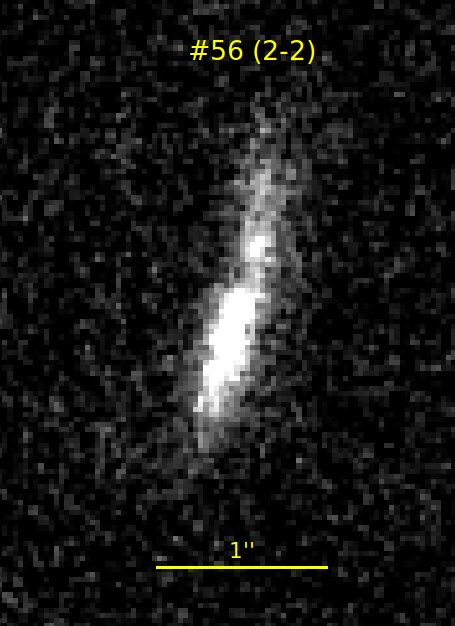}  
\end{center}
\end{figure*}

\end{appendix}

\end{document}